\documentclass[%
 reprint,
 superscriptaddress,
 amsmath,amssymb,
 aps,
 nonatbib,
]{revtex4-2}

\makeatletter

\let\textcite\relax
\let\citet\relax
\let\citep\relax

\expandafter\let\csname ver@natbib.sty\endcsname\relax
\makeatother
\usepackage[style=numeric,backend=biber, sorting=none]{biblatex}
\makeatletter
\def\blx@err@patch#1{%
  \PackageWarning{biblatex}{Patching #1 failed; continuing anyway}%
}
\makeatother

\usepackage[english]{babel}
\usepackage{url}
\usepackage{graphicx}
\usepackage{dcolumn}
\usepackage{bm}
\usepackage{comment}
\usepackage{amsmath,amssymb}
\usepackage{tikz}
\usepackage[colorlinks=true,bookmarks=false,citecolor=blue,urlcolor=blue]{hyperref} 
\usepackage{physics}

\usepackage[autostyle]{csquotes}
\MakeOuterQuote{"}

\usepackage{xcolor}
\definecolor{hamza-color}{RGB}{0,80,250}

\newcommand{\tensorepsilon}{\overline{\overline{\varepsilon}}}

\makeatletter
\newenvironment{supplement}{%
  \setcounter{secnumdepth}{3}%
  \setcounter{section}{0}%
  \setcounter{figure}{0}%
  \setcounter{table}{0}%
  \def\@seccntformat##1{\csname the##1\endcsname:\ }%
}{%
  \def\@seccntformat##1{\csname the##1\endcsname.\ }%
}
\makeatother

\newcommand{\SMsec}[1]{Sec.~\ref{#1}}

\begin{document}
\renewcommand{\bibliography}[1]{}







\title{Full-Stack High-Volume Quantum Networking Architecture based on Photonic-Integrated Tin Vacancy Centers in Diamond}

\author{Hamza Raniwala}
\thanks{These authors contributed equally}
\email{raniwala@mit.edu}
\affiliation{Department of Electrical Engineering and Computer Science, Massachusetts Institute of Technology, Cambridge, MA 02139, USA}\affiliation{Research Laboratory of Electronics, Massachusetts Institute of Technology, Cambridge, MA 02139, USA}

\author{Ian Christen}\thanks{These authors contributed equally}
\affiliation{Department of Electrical Engineering and Computer Science, Massachusetts Institute of Technology, Cambridge, MA 02139, USA}\affiliation{Research Laboratory of Electronics, Massachusetts Institute of Technology, Cambridge, MA 02139, USA}

\author{Helaman Flores}\thanks{These authors contributed equally}
\affiliation{Department of Electrical Engineering and Computer Science, Massachusetts Institute of Technology, Cambridge, MA 02139, USA}\affiliation{Research Laboratory of Electronics, Massachusetts Institute of Technology, Cambridge, MA 02139, USA}

\author{David Starling} 
\affiliation{Lincoln Laboratory, Massachusetts Institute of Technology, Lexington, Massachusetts 02421, USA}

\author{Ryan Murphy} 
\affiliation{Lincoln Laboratory, Massachusetts Institute of Technology, Lexington, Massachusetts 02421, USA}

\author{Eric Bersin}
\affiliation{Lincoln Laboratory, Massachusetts Institute of Technology, Lexington, Massachusetts 02421, USA}

\author{Kevin Chen}
\affiliation{Department of Electrical Engineering and Computer Science, Massachusetts Institute of Technology, Cambridge, MA 02139, USA}\affiliation{Research Laboratory of Electronics, Massachusetts Institute of Technology, Cambridge, MA 02139, USA}

\author{Marc Davis}
\affiliation{Department of Electrical Engineering and Computer Science, Massachusetts Institute of Technology, Cambridge, MA 02139, USA}\affiliation{Research Laboratory of Electronics, Massachusetts Institute of Technology, Cambridge, MA 02139, USA}

\author{Maxim Sirotin}
\affiliation{Department of Electrical Engineering and Computer Science, Massachusetts Institute of Technology, Cambridge, MA 02139, USA}\affiliation{Research Laboratory of Electronics, Massachusetts Institute of Technology, Cambridge, MA 02139, USA}

\author{Mahmoud Jalali Mehrabad} 
\affiliation{Department of Electrical Engineering and Computer Science, Massachusetts Institute of Technology, Cambridge, MA 02139, USA}\affiliation{Research Laboratory of Electronics, Massachusetts Institute of Technology, Cambridge, MA 02139, USA}

\author{Ethan G. Arnault} 
\affiliation{%
Department of Electrical Engineering and Computer Science, Syracuse University, Syracuse, NY 13244, USA
}
\affiliation{%
Institute for Quantum and Information Sciences, Syracuse University, Syracuse, NY 13244, USA
}%
\affiliation{%
Department of Physics, Syracuse University, Syracuse, NY 13244, USA
}%

\author{Matthew E. Trusheim} 
\affiliation{Department of Electrical Engineering and Computer Science, Massachusetts Institute of Technology, Cambridge, MA 02139, USA}\affiliation{Research Laboratory of Electronics, Massachusetts Institute of Technology, Cambridge, MA 02139, USA}

\author{P. B. Dixon} 
\affiliation{Lincoln Laboratory, Massachusetts Institute of Technology, Lexington, Massachusetts 02421, USA}

\author{Dirk R. Englund} 
\email{englund@mit.edu}
\affiliation{Department of Electrical Engineering and Computer Science, Massachusetts Institute of Technology, Cambridge, MA 02139, USA}\affiliation{Research Laboratory of Electronics, Massachusetts Institute of Technology, Cambridge, MA 02139, USA}

\date{\today}
\begin{refsection}
\begin{abstract}

Solid state quantum emitters~\cite{donnelly2026large,edlbauer202511} are a leading platform for photonic quantum networking with memory nodes. However, the inhomogeneous distribution of quantum emitters, as well as several environmental factors (i.e. strain and electric fields) spread the frequency spectrum of the qubits, making them distinguishable and therefore not a reliable resource for distributed quantum entanglement. In this paper, we demonstrate a full-stack approach to integrating nearly indistinguishable tin vacancy (SnV$^-$) quantum emitters on a frequency-tunable photonic interposer that overcomes the native distribution and static variation of quantum emitters for an indistinguishable photonic quantum networking platform. We demonstrate a silicon nitride-on-insulator photonic integrated circuit (PIC) with accompanying multiphysics digital twin (MPhDT) that guides discovery of SnV$^-$ strain-tuning parameters and informs construction of a multi-channel quantum repeater node. On this node, we achieve the first simultaneous demonstration of spectral tuning of the zero phonon line (ZPL) at GHz scale; coherent electron spin control with gate times of $<80$ ns; strongly- and weakly-coupled nuclear spin detection; and commercial fiber array-coupled readout of a SnV$^-$ center. Finally, we propose and simulate improvements to the architecture that achieve 99.96\% connectivity of $ N \sim 1000$ emitters spanning the inhomogeneous distribution of SnV$^-$ centers in strained diamond, where distributed quantum entanglement may be realized.






\end{abstract}

\maketitle

\section{Introduction}

\begin{figure*}
    \centering
    \includegraphics[width=\textwidth]{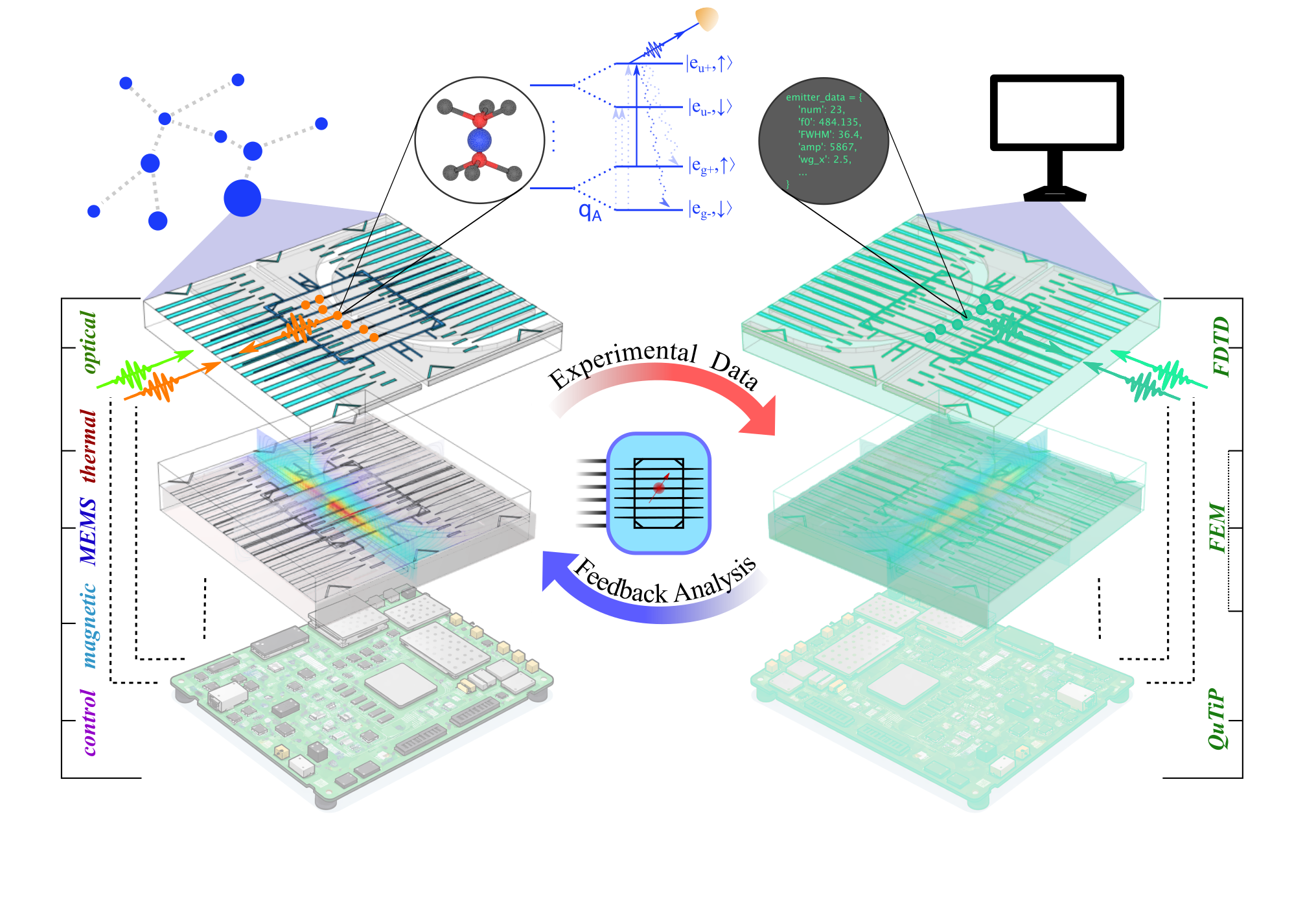}
    \caption{Engineering a full-stack blueprint of a QR-PIC. The left hand side of the figure diagrammatically depicts real-world components of a quantum network, including an entangled network (blue dots) where each node consists of a QR-PIC (left top) housing SnV$^-$ quantum emitters (inset). The QR-PIC exhibits multiphysics phenomena, including optical interfacing with SnV$^-$ emitters in nanophotonic waveguides (left second), thermal (left third) and controllable electromechanical (left fourth) strain effects perturbing quantum resources on-chip, magnetic field extents (left fifth) of coherent electromagnetic drives of SnV$^-$ electron spin qubits, and real-world electronics governed by an RFSoC (left bottom). These components are captured in a multiphysics digital twin (MPhDT) (right), where emitters are recorded in a sample description database (right top), multiphysics phenomena are captured with FDTD and FEM simulation software (right middle), and quantum control is modeled in Python QuTiP (right bottom). The feedback between real-world and digital components creates a data-analysis feedback loop that crafts a full-stack quantum repeater.}
    \label{fig:overview}
\end{figure*}

Substantial advances in quantum networking over the last decade have largely been driven by the entanglement of single photons. Several platforms have been developed to achieve single photon transfer and entanglement of quantum information, including probabilistic spontaneous parametric down-conversion based systems\cite{lu2009experimental, arahira20121}, single photon emitters such as quantum dots\cite{yu2023telecom, chan2025practical}, and recently gases of neutral atoms \cite{liu2024creation}. These single-photon quantum networking platforms have paved the way for single-photon entanglement of remote Bell pairs\cite{knaut2024entanglement, stas2026entanglement, ruskuc2025multiplexed} and quantum key distribution\cite{bhaskar2020experimental}.

Among single-photon emitters, quantum emitters in diamond known as color centers--including nitrogen vacancy centers and Group IV (silicon, germanium, tin, lead)--have demonstrated results in memory-assisted quantum networking~\cite{Bersin2024}. In particular, the SnV$^-$ emitter has recently enabled broadband-waveguide-coupled emitter entanglement, marking it as a promising candidate for large-scale quantum networks \cite{waas2026remote}. Adjacent efforts have been made to scale the optical addressability of quantum emitters using heterogeneous architectures such as photonic integrated circuits and CMOS backplanes \cite{wan2020large, li2024heterogeneous}. 

However, translating these demonstrations to large entangled clusters of emitters poses a major obstacle: the local properties of individual solid-state emitters are not uniform and not fully known in advance, resulting in three interrelated challenges that limit the scalability of color center-based quantum photonic systems.
\begin{itemize}
    \item \textbf{(C1) Unknown global emitter parameters.} Predictively designing strain tuning for a given device requires knowledge of the SnV$^-$ strain susceptibility parameters $t_{\perp}$ and $t_{||}$, the transverse and axial ZPL strain susceptibilities per unit strain. The axial parameter $t_{||}$ has been reported for an individual emitter of uncertain absolute position \cite{clark2024nanoelectromechanical}, resulting in large measurement uncertainties that propagate into errors in predicted tuning ranges. The parallel parameter $t_{||}$ has not yet been reported. Without reliable values of $t_{\perp}$ and $t_{||}$, strain tuning cannot be designed from first principles and device geometries cannot be optimized.
    \item \textbf{(C2) ZPL inhomogeneity.} Variations in local crystal strain shift the ZPL of each SnV$^-$ by amounts that can exceed the inhomogeneous linewidth $\Gamma_{inh}$ in fabricated devices, preventing photon indistinguishability and therefore most entanglement protocols without active frequency control \cite{iwasaki2015germanium, sutula2023large, eremchev2021microscopic}.
    \item \textbf{(C3) Scalable coherent control across multiple channels.} Translating spectral indistinguishability into useful entanglement requires coherent spin control on each optically addressable qubit. Achieving this across many channels simultaneously has not been demonstrated in a multi-channel PIC platform.
\end{itemize}
Addressing these challenges in isolation has been the subject of substantial recent work: electromechanical strain tuning of SnV$^-$ centers has been demonstrated in nanophotonic waveguides \cite{clark2024nanoelectromechanical, brevoord2025large}, and coherent spin control of SnV$^-$ has been achieved in bulk \cite{rosenthal2023microwave, rosenthal2024single, karapatzakis2024microwave, resch2026high} and in diamond membranes via strain-enabled microwave driving \cite{guo2023microwave}. However, a unified methodology capable of characterizing the device, extracting emitter parameters needed to design strain tuning, and deploying coherent control--all within an integrated platform--has yet been lacking.
Here, we introduce a multi-layered framework of hardware, firmware, and software to overcome these challenges. We develop a commercial fiber array-coupled silicon nitride-on-insulator photonic integrated circuit with embedded microwave lines and DC electrodes to control SnV$^-$ vacancies heterogeneously integrated via quantum microchiplets (QMCs) as a fully functional quantum repeater-photonic integrated circuit (QR-PIC) architecture. In parallel, we construct a multiphysics digital twin (MPhDT) of the QR-PIC. The MPhDT is a per-sample computational replica of a physical QR-PIC, constructed from the actual pick-and-stamp (PnS) geometry of each heterogeneously integrated QMC using computer vision and fleshed out with finite-element method (FEM) and finite-difference time domain (FDTD) simulations of thermo-electromechanical strain, optical coupling, and microwave field distributions. Critically, we show that when the MPhDT's simulated strain environment is coupled to experimental ZPL shift data via an expectation-maximization (EM) algorithm, the system simultaneously resolves global strain susceptibility parameters as well as the latent per-emitter position and orientation--an inverse problem that cannot be fully solved by analyzing emitters in isolation. 

Using the hardware-digital twin stack as our analytical backbone, we demonstrate three interconnected results. First, we extract SnV$^-$ ZPL strain susceptibilities from an ensemble of 27 SnV$^-$ emitters distributed across seven QMCs, providing the first complete characterization of the SnV$^-$ ZPL strain susceptibility relevant to integrated photonic devices to address C1. Second, we use these parameters to predictively guide electromechanical tuning of a SnV$^-$ ZPL in a QR-PIC coupled via fiber array in a Bluefors cryostat, demonstrating C2 via MPhDT-informed spectral control that achieves approximately 30 MHz-error predictive accuracy (around the lifetime limited linewidth of SnV$^-$ emitters \cite{trusheim2020transform}) up to 100 V. Third, we deploy coherent microwave control of the electron spin qubit of the strain-tunable emitter to address C3, establishing--to our knowledge--the first simultaneous demonstration of spectral tuning, spin-photon readout, and nuclear spin control in a QR-PIC coupled to a commercial single-mode fiber array. Finally, using the extracted parameters as inputs, the MPhDT predicts that MEMS-actuated QMCs deployed in the QR-PIC enable spectrally connected ensembles of $N > 100$ qubits per device, providing a concrete and quantitatively grounded roadmap for scaling quantum repeater nodes.

\begin{figure*}[t]
    \centering
    \includegraphics[width=\textwidth]{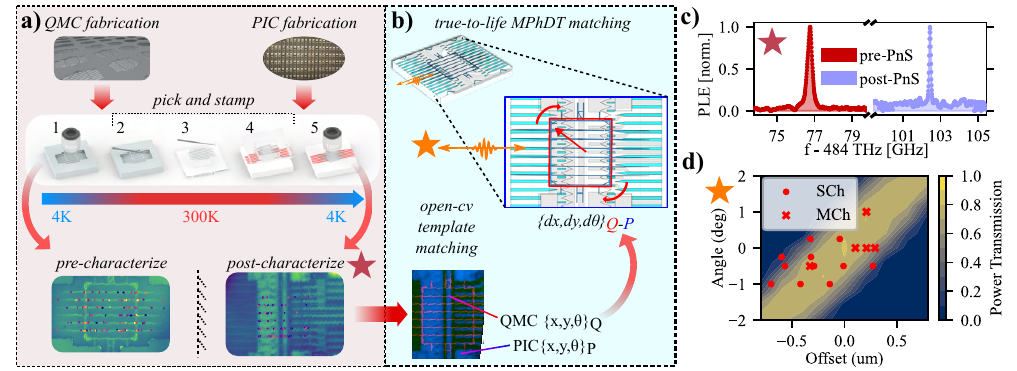}
    \caption{Development of physical and digital quantum repeater. (a) real-world development of the QR-PIC, starting from QMC and PIC fabrication components which are combined in a heterogeneous pick-and-stamp procedure aligning a QMC to the PIC photonic integrated circuit. Emitters in the QMC are characterized using a widefield PLE (w-PLE) technique \cite{sutula2023large} to identify emitters pre- and post-PnS (red star). (b) digital twin sample construction, including a FEM and FDTD (not shown) software recreation from the sample planar geometry file. QMC (Q)-PIC (P) offsets are measured using an OpenCV template matching algorithm, and these offsets $(dx, dy, d\theta)_{Q-P}$ are passed to simulation to evaluate metrics such as expected waveguide-to-waveguide coupling efficiency (gold star) and thermo-electromechanical effects. (c) representative PLE of an emitter pre- and post-PnS, with $>20$ GHz frequency shifts observed. (d) digital evaluation of power transmission from the QMC waveguide to PIC channel, with scatter plot of single-channel (SCh) and multi-channel (MCh) QMC-coupled samples and their respective simulated waveguide-to-waveguide power transmission. The underlying contour plot sweeps diamond waveguide offset and angle against diamond waveguide-to-PIC power transmission.}
    \label{fig:sample_construction}
\end{figure*}
\section{High-Yield Manufacturing of QR-PICs with integrated SnV$^-$ QMCs}

The physical components of our quantum networking device are a silicon nitride-on-oxide photonic integrated circuit (PIC) and a diamond nanophotonic quantum microchiplet (QMC). 
The PIC, designed at MIT Lincoln Laboratory \cite{starling2023fully}, hosts one or more diamond nanophotonic quantum microchiplets (QMCs) in lithographically defined sockets and provides low-loss optical routing to commercial single-mode fiber arrays, on-chip microwave transmission lines for spin control, and buried DC electrodes for electromechanical strain actuation of QMC waveguides. The QMC is a suspended chassis of diamond nanobeam photonic waveguides that houses SnV$^-$ centers implanted by ion implantation and annealing \cite{chen2024scalable}. Together, the QR-PIC defines a complete spin-photon interface where photons emitted by SnV$^-$ centers into QMC waveguides are evanescently coupled to SiN PIC channels and routed to optical fiber; microwave fields from the on-chip transmission line drive electron spin transitions; and DC-biased electrodes strain-shift emitter ZPLs in a given QMC channel.

These components are combined in a heterogeneous integration process for which we build a digital twin, outlined in Fig. \ref{fig:sample_construction}. QMC fabrication is done on bulk diamond samples using the fabrication techniques outlined in \cite{chen2023protocols, chen2024scalable}, while the PIC fabrication is completed in a foundry-compatible Lincoln Laboratory process \cite{starling2023fully}. In the quantum microchiplet, we embed SnV$^-$ centers using ion implantation and annealing of bulk diamond, followed by electron beam lithography of nanophotonic waveguides constricted in a mechanically stable chassis. Once these components are completed, we proceed with a novel pick-and-stamp procedure consisting of a high-yield PDMS mass transfer of QMCs into pre-defined QMC sockets in the PIC, shown in Fig. \ref{fig:sample_construction}a.

To integrate our measured data with our MPhDT for performance validation, we can recreate the QR-PIC in FEM and FDTD digital environments by importing CAD files for both the PIC and QMC and using computer vision techniques (Fig. \ref{fig:sample_construction}b) \cite{OpenCV_TemplateMatching}. Using OpenCV template matching, we identify both the PIC and QMC locations in a $\sim50$ $\mu$m field of view with $\sim100$ nm precision--limited by white light image resolution--to identify the translational-rotational offset $(dx, dy, d\theta)_{Q}$ of the QMC origin with respect to the PIC. We pass this information to FEM and FDTD simulation suites, primarily COMSOL Multiphysics and Tidy3D \cite{Tidy3D}, to recreate seven PnS MPhDT samples in the digital domain. From there, we can evaluate metrics such as the waveguide-to-waveguide power coupling between QMC and SiN PIC (Fig. \ref{fig:sample_construction}d) and the local strain environment of our emitters (next section). Among all coupled waveguides, we find a mean coupling of $0.61 \pm 0.22$, capturing $75\% \pm 27\%$ of the maximum coupling (81\%), with the mean performance diminished due to a pitch mismatch between QMC and PIC waveguides. However, when removing waveguides mismatched due to pitch, we find a mean coupling of $0.74 \pm 0.05$ ($91\% \pm 6\%$ of the maximum). Thus, our MPhDT verifies the relative performance of our heterogeneous integration technique, affirming its scalability at $>90\%$.

Documenting SnV$^-$ within a QMC is critical to this procedure. Individual SnV$^-$ centers differ with respect to one another in crystal lattice orientation $\vec{o} \in \{[\pm1, \pm1, 2]\}$, while the waveguide crystal axis is $[1,1,0]$, yielding two orientation classes of emitters: transverse ($[1,-1,2], [-1,1,2]$) and axial ($[1,1,2], [-1,-1,2]$), with axial emitters' dipole moments better coupled to the TE waveguide mode (see Supplemental). Additionally, local crystal strain $\tensorepsilon$ influences the energy level structure of each SnV$^-$, shifting and stretching the levels depicted in Fig.~\ref{fig:overview} to result in unique zero phonon line frequencies $f_{ZPL,i}$ and qubit frequencies $f_i$ for each SnV$^-$ $i$. In particular, the zero phonon line $f_{ZPL,i}$ is of interest in this work, as aligning quantum emitter frequencies is critical to photonic indistinguishability of emitters and therefore most entanglement procedures \cite{ollivier2021hong}. Under local strain, the ZPL location is given by
\begin{equation}
    f_{ZPL,i} = f_{0} + t_{\perp}(\varepsilon_{xx}+\varepsilon_{yy}) + t_{||}(\varepsilon_{zz}).
\end{equation}
Here, $t_{\perp} = t_{\perp,e} - t_{\perp,g}$ is the transverse strain susceptibility difference between the excited orbital state and the ground state of the SnV$^-$, while $t_{||} = t_{||,e} - t_{||,g}$ is the axial variant \cite{hepp2014electronic, meesala2018strain}. We importantly note that $f_{ZPL,i}$ can be further broken down as
\begin{equation}
    f_{ZPL,i} = f_{pre,i} + \Delta ZPL_i,
\end{equation}
where $\Delta ZPL_i$ is the zero phonon line shift as a result of both strain induced by the PnS process as well as active strain tuning on the QR-PIC (Fig. \ref{fig:sample_construction}c). Therefore, we document both $f_{pre,i}$ and $\Delta ZPL_i$ in the PnS process by characterizing the QMC under widefield photoluminescence excitation (PLE) (see Supplement for details). This information becomes critical for MPhDT development, which we will outline in the next section.

\begin{figure}
    \centering
    \includegraphics[width=\columnwidth]{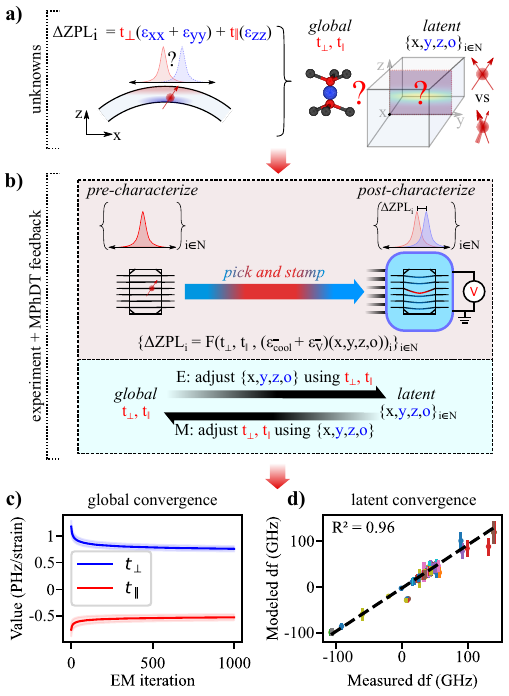}
    \caption{Experiment-analysis feedback revealing unknown global and local latent variables of SnV$^-$ quantum resources. (a) illustration of on-chip strain in a diamond waveguide generating frequency shifts in the ZPL of a SnV$^-$ post-PnS and upon applied MEMS voltage. The ZPL shift is difficult to predict \textit{a priori} due to the globally uncertain and underreported variables of $t_{perp}$ and $t_{par}$, as well as the latent variables of each emitter (waveguide cross-sectional location, orientation). (b) Expectation-maximization (EM) algorithm run on seven experimental QMCs of the PnS procedure and applied MEMS voltage on test chips. The ZPL shift of emitters in each sample are observed and recorded, and the collective data is inserted in a MPhDT-informed algorithm that takes simulated cross-sectional waveguide strain at each emitter location with the experimentally observed ZPL shifts and iteratively re-tunes the global variables \{$t_{perp}$, $t_{par}$\} and re-fits the expected \{(y,z,o)\}$_{i \in N}$ of each emitter $i$. (cd) convergence of the EM algorithm on both global (c) and latent (d) fronts across 100 Monte Carlo algorithm runs.}
    \label{fig:em_maximization}
\end{figure}

\begin{figure*}[t]
    \centering
    \includegraphics[width=\textwidth]{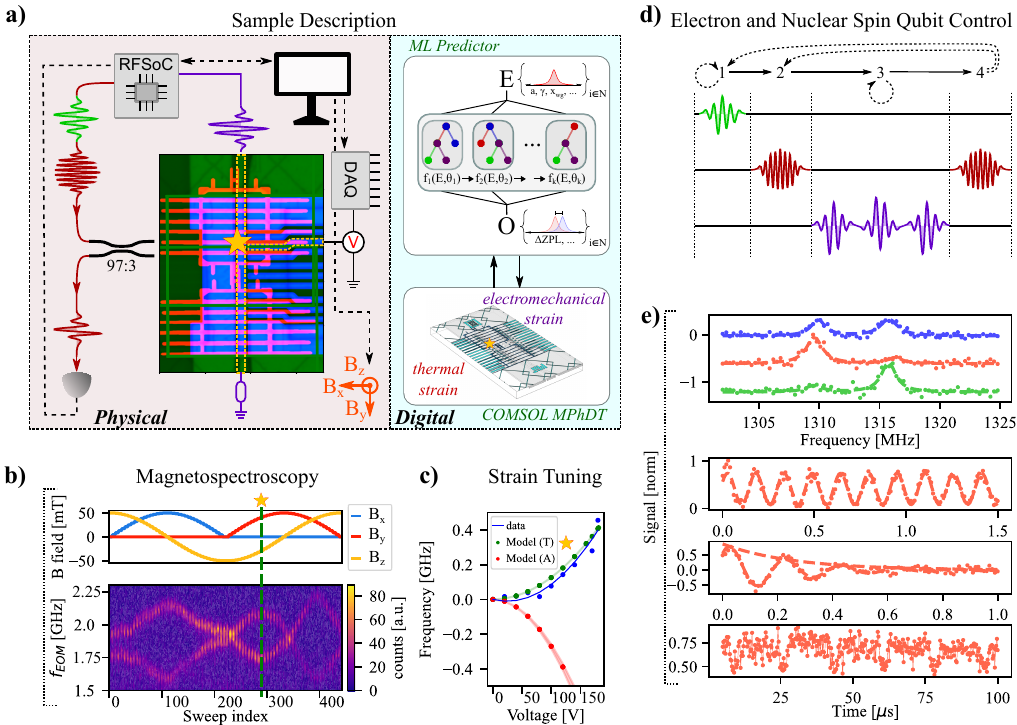}
    \caption{Robust control and model understanding of a sample emitter in a Bluefors cryostat. (a) sample constructed of a QR-PIC with two microchiplets stamped in the socket, with the targeted emitter pinpointed with a gold star. The emitter is modeled with an XGBoost model interacting with the sample MPhDT to provide expected spectral tuning parameters. (b) spectral characterization of the targeted emitter using magnetospectroscopy and (c) MEMS voltage-tuned strain spectroscopy. The digital twin model captures, with root-mean-squared error of $30$ MHz up to 100 V, the transverse orientation and ZPL shift from zero voltage of the emitter. (d) knowledge of the emitter location near the microwave line permits robust electron spin and nuclear spin qubit control via an RFSoC programmed state machine (top) applying a coherent pulse sequence to the emitter (bottom, simplified). (e) results of pulse sequence control on the targeted emitter. (top) ODMR reveals a nuclear-coupled hyperfine spin structure (blue), which can be initialized in nuclear spin-up (orange) and spin-down (green) states using variable optical polarization sequences (d,2). Rabi (second), Ramsey (third), and XY8 (bottom) experimental results, operated using different microwave gate sequences (d,3).}
    \label{fig:prediction_and_control}
\end{figure*}

\section{MPhDT Feedback on Unknown Emitter Parameters}
In large scale quantum networking systems, knowledge of the global parameters of the chosen quantum emitter is vital. However, current knowledge of the SnV$^-$ strain susceptibility parameters is limited to analysis of individual emitters under uncertain absolute locations \cite{clark2024nanoelectromechanical, brevoord2025large}. This can yield large uncertainties in the tunability of emitters \textit{in situ}, the frequency of electron spin qubit transitions under magnetic field, etc. In this work, we demonstrate the usage of a MPhDT to elucidate the $t_{\perp}$ and $t_{||}$ parameters highlighted above and in Fig. \ref{fig:em_maximization}a. We surmise that $\Delta ZPL_i$ is likely dominated by crystal strain perturbations to the waveguide post-integration, following the literature on quantum emitter zero-phonon line susceptibilities \cite{knauer2020situ}. The strain tensor $\tensorepsilon$ producing this effect may result from thermal compression due to mismatched thermal expansion coefficients of the QMC and PIC layers, and additionally from electromechanical strain as a result of DC bias electrodes buried under the PIC SiN layer (see Supplement). Therefore, we extract the cross-sectional "strain map" $\tensorepsilon(y,z)$ in the waveguide at the location of each emitter identified in both pre- and post-PnS w-PLE data. We compile the observed $\Delta ZPL_i$ and simulated strain maps to an expectation-maximization (EM) algorithm (Fig. \ref{fig:em_maximization}b). The algorithm alternates for 1000 iterations between two steps:
\begin{itemize}
\item \textbf{E-step:} Update responsibility weight of each $(o,y,z)$ assignment for emitter $i$ based on agreement with the measured $\Delta ZPL_i$. This assigns higher probability to $(o,y,z)$ configurations that better predict the observed ZPL shift.
\item \textbf{M-step:} Update $(t_{\perp}, t_{||})$ by minimizing the responsibility-weighted squared residual. This drives the global parameters toward values that best explain the entire ensemble under the current emitter position estimates.
\end{itemize}

We run the above algorithm for 36 different clusters of 100 Monte Carlo runs centered around a grid of initial guesses of $t_{\perp}$ and $t_{||}$. Each cluster of Monte Carlo runs converges to a different $t_{\perp}$ and $t_{||}$ and cross-sectional position distribution, and we break the tie by highest R$^2$ value between measured $\Delta ZPL_i$ and simulated ZPL shift as well as emitter location distribution profile match to TRIM. Fig. \ref{fig:em_maximization}c-d shows the convergence of $t_{\perp}\rightarrow0.757 \pm 0.049\text{ PHz/strain}$ and $t_{||} \rightarrow -0.520 \pm 0.056\text{ PHz/strain}$. The extracted $t_{||}$ is comparable to the prior literature value of approximately $-0.460$ PHz/strain \cite{clark2024nanoelectromechanical}, providing a useful cross-validation to our methodology. The value of $t_{\perp}$ is, to our knowledge, the first reported measurement of the SnV$^-$ transverse strain susceptibility in an integrated photonic device--a parameter that directly impacts the design of electromechanical tuning electrodes and prediction of ZPL strain tuning curves vs applied voltage. The R$^2$ of the ensemble fit is 0.96, quantifying the quality of the MPhDT strain model.


\section{Implementation of Tuning and Control on a QR-PIC in a Dilution Fridge}
\begin{figure*}[t]
    \centering
    \includegraphics[width=\textwidth]{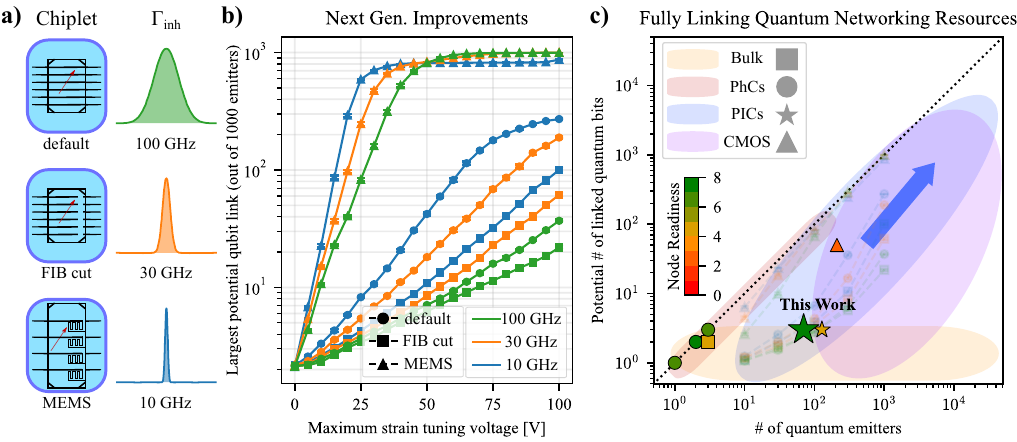}
    \caption{The MPhDT enables smarter design of next-generation QR-PICs. (a) exploring the QMC sample space by varying an input QMC to a disjoint set union (DSU) algorithm, where unions are made between emitters when one emitter's ZPL strain tuning curve overlaps with another emitter's curve at any voltage of both curves (assuming local tuning shown in this work). Chiplet types are varied from the default and FIB-cut waveguide QMCs (both in this work) to MEMS-designed single-sided cantilever waveguides. The inhomogeneous linewidth $\Gamma_{inh}$ of SnV$^-$ emitters is also varied from 100 GHz (observed in this work) to 10 GHz (4 GHz observed in \cite{narita2023multiple}), providing a 2D matrix of targeted QMC improvements. (b) size of largest potential qubit link as a function of the next generation improvement matrix, showing that large, fully resourced qubit links are possible with lower PIC voltage requirements dependent on QMC improvements. (c) roadmap of the QR-PIC blueprint \textit{vis a vis} other quantum networking platforms, including bulk/confocal optics \cite{pompili2021realization, stolk2024metropolitan}, photonic crystal spin-photon interfaces \cite{knaut2024entanglement,ruskuc2025multiplexed}, other PIC work \cite{wan2020large}, and CMOS architectures \cite{li2024heterogeneous}. The "Node Readiness" score of each literature result consists of a sum of 0, 1, or 2 for each of spectral tuning, spin control, fiber coupling, and entanglement demonstration; a score of 2 represents a demonstrated quantum repeater feature; 1 represents an achievable but not demonstrated feature; and 0 represents a feature unachievable without device overhaul.}
    \label{fig:system_scaling}
\end{figure*}

With MPhDT and tuned strain susceptibilities in hand, we deploy a QR-PIC in a Bluefors cryostat operating at 1K to demonstrate the critical components for memory-based quantum networking: spectral tuning, spin control, and fiber-based readout. The device under test is shown in Fig. \ref{fig:prediction_and_control}. We detect $\sim140$ emitters between $483.86$ THz to $484.30$ THz across four diamond waveguides coupled QR-PIC channels using w-PLE at 4 Kelvin. However, upon cooling in the dilution refrigerator and switching from free space to fiber array coupling, we are only able to distinguish $\sim70$ emitters across all channels. While the exact cause for this discrepancy is unknown, we may attribute it to limited pump power or collection efficiency through the PIC, overlapping emitter peaks, or emitter ZPLs falling outside the fiber-measured frequency range ($483.95$ THz to $484.12$ THz).

Naively, one might expect that the low number of emitter $\Delta ZPL_i$ data points ($N = 189$ after aggressive filtering of noisy data) across 30 emitters within our post-PnS w-PLE data would not lend itself to complex machine learning models. However, we deploy an Extreme Gradient Boosting (XGBoost) supervised learning regression model (titled "ML Predictor" in Fig. \ref{fig:prediction_and_control}a) to guess the tuning of an emitter identified under pre-PnS w-PLE \cite{chen2016xgboost}. We experimentally validate the ML prediction by using a magnetic field sweep to determine the transverse nature of the targeted SnV$^-$ (Fig.~\ref{fig:prediction_and_control}b). Without post-PnS w-PLE information on the emitter, we still find remarkable agreement up to 100 V of DC bias with XGBoost model-predicted transverse electromechanical spectral tuning (Fig.~\ref{fig:prediction_and_control}c). This also affirms our extracted $t_{\perp}, t_{||} $ values from the previous section.

We proceed with coherent control of the SnV$^-$ electron spin using the on-PIC microwave transmission line under the QMC (vertical yellow dotted strip in Fig. \ref{fig:prediction_and_control}a). Using a custom waveform generation package running on a Xilinx ZCU111 RFSoC titled \textit{pulseseq} (see \SMsec{sm:pulseseq}), we operate a state machine (Fig. \ref{fig:prediction_and_control}d) 
that implements a four-state protocol: (1) charge-state confirmation with a 515 nm green repump pulse; (2) spin polarization into $\ket{e_{g-},\downarrow}$ with a resonant 484.05 THz (619.34 nm) laser; (3) application of a desired microwave pulse sequence; and (4) resonant optical spin readout (Fig.~\ref{fig:prediction_and_control}d). The results of four canonical pulse experiments are shown in Fig.~\ref{fig:prediction_and_control}e:
\begin{itemize}
    \item \textbf{ODMR} reveals a hyperfine splitting of the electron spin resonance, indicating the presence of a strongly coupled $^{13}\text{C}$ nuclear spin. Optical polarization sequences initialize the nuclear spin into $\ket{\uparrow}$ and $\ket{\downarrow}$ states, demonstrating electron-nuclear spin register control.
    \item \textbf{Rabi oscillations} at $\Omega = 6.49 \pm 0.01$ MHz yield a $\pi$-pulse time of $76.97\pm0.11$ ns at 16 dBm input power.
    \item \textbf{Ramsey interferometry} measures a pure dephasing time of $T_{2}^* = 256 \pm 2$ ns, a value consistent with natural-abundance $^{13}\text{C}$ bath decoherence and representing the expected floor for this material system. Importantly, this timescale permits spin decoupling with $\pi$-pulse time around 5x shorter than the pure dephasing, permitting dynamical decoupling experiments.
    \item \textbf{XY8 dynamical decoupling} reveals collapse and revival signatures consistent with four weakly coupled $^{13}\text{C}$ nuclear spins near the $^{13}\text{C}$ Larmor period under 0.05 T, identified by linear spacing at $k\text{th}$ order harmonics (see \SMsec{sm:spin-echo}). This confirms the potential of the QR-PIC platform to leverage nuclear spins as long-lived quantum memory registers.
\end{itemize}
To our knowledge, this is the first simultaneous demonstration of multi-channel spectral tuning, electron spin control, nuclear spin control, and fiber-coupled readout in a multi-channel PIC platform using commercial single-mode fiber arrays.

\section{Scaling of Future QR-PICs}

Having verified that our MPhDT encapsulates the performance of the SnV$^-$ quantum resources in our QR-PIC, we proceed with evaluating the major opportunities to scale to exponentially larger quantum networking platforms. We evaluate our device by randomly generating $N \in \{10, 30, 100, 300, 1000\}$ emitters from an inhomogeneous ZPL distribution with inhomogeneous linewidth $\Gamma_{inh}$. We randomly assign each emitter to a $(x,y,z)$ location in a QMC, and we simulate the $\Delta ZPL_i$ experienced by that emitter upon PnS and deployment in a cryostat. Finally, we simulate the electromechanical strain tuning curve $\Delta ZPL_{app, i}(V) = t_{\perp}(\varepsilon_{xx}(V) + \varepsilon_{yy}(V)) + t_{||}\varepsilon_{zz}(V)$ for all emitters. This is a sum total of the spectral information of a MPhDT-informed digital emitter ensemble of size $N$. We determine potential emitter links by checking when an emitter ZPL curve $f_{ZPL,i}(V) = f_{ZPL,i} + \Delta ZPL_i + \Delta ZPL_{app, i}(V)$ overlaps with another curve $f_{ZPL,j}$; if the curves overlap in frequency space, the emitters are considered potentially linked \cite{li2024heterogeneous}. Using a disjoint set union (DSU) algorithm, we determine the largest potential qubit link for each digital emitter ensemble. We repeat this for a matrix of potential improvements. The first matrix axis is $\Gamma_{inh} \in \{100, 30, 10\}$ GHz, replicating a progressively narrower inhomogeneous $f_{ZPL}$ distribution following state-of-the-art emitter implantation and annealing techniques. The second axis is QMC design--whereas in this work we utilized nanophotonic waveguides clamped in a chassis or FIB cut to generate unique strain environments and tuning curves for various emitters. 

Fig.~\ref{fig:system_scaling}b shows the largest spectrally connected ensemble as a function of N for each QMC-$\Gamma_{inh}$ configuration. Two results stand out. First, we find that the transition from basic and/or FIB-cut QMC waveguides to MEMS structures provides a one-to-two order of magnitude increase in connected ensemble size at fixed $\Gamma_{inh}$--a larger gain than reducing $\Gamma_{inh}$ alone. Second, the MEMS configurations irrespective of $\Gamma_{inh}$ result in nearly fully resourced emitter chains with $N > 300$ connected emitters (saturating at the number of emitters in the simulation) for large emitter count, with decreasing voltage required to achieve the chain the more emitters simulated. For example, at $N=300$ emitters, the smallest emitter chain connectivity $C = N_{chain}/N$ across $\Gamma_{inh}$ variations is $82.02\pm0.25\%$; at $N=1000$, this rises to at least $86.98\pm0.66$\% ($\Gamma_{inh} = 10$ GHz) and as high as $99.96\pm0.01$\% ($\Gamma_{inh} = 30$ GHz). This makes MEMS QMC development with high emitter density the highest priority near-term improvement to the QR-PIC platform.

We note explicitly that spectral connectivity is a necessary but not sufficient condition for entanglement. To generate useful heralded Bell pairs from overlapping emitters, we additionally require linewidth matching (Hong-Ou-Mandel visibility depends on both frequency and linewidth; see \SMsec{sm:strain-tuning} for a treatise of emitters in the sample in this work), phase stability over the measurement window, spin echo lifetimes longer than the Bell Pair entanglement procedure, typically 10's of $\mu\text{s}$; similar dipole orientation projections into the collection mode; and adequate collection efficiency through the full optical path. An optical link budget of known losses in the current QR-PIC is considered in \SMsec{sm:FDTD}.

Fig.~\ref{fig:system_scaling}c benchmarks the QR-PIC architecture against the current literature across four metrics: spectral tuning capability, coherent spin control, fiber coupling, and entanglement demonstration. Each literature result is assigned a "Node Readiness" inspired by NASA's Technology Readiness Level (TRL) to assess sample deployability. We assign each result a Node Readiness score outside the context of the \textit{scalability} metrics of the number of emitters vs the potential number of linked quantum bits. This allows us to conveniently visualize the progress of each platform towards scalable quantum networking. Our platform simultaneously achieves demonstrated or demonstrable status across all four metrics for the first time in an SnV$^-$-based PIC architecture--a noteworthy milestone as the combination of high Debye-Waller factor SnV$^-$ emitters \cite{gorlitz2020spectroscopic, trusheim2020transform}  with broadband photonic waveguides overcomes the limitation of cavity-to-cavity inhomogeneity in photonic crystal-based approaches to spin-photon interfaces \cite{kuruma2021coupling, rugar2021quantum, lee2026quantum, ruskuc2025multiplexed, chen2016xgboost}. The MPhDT-guided design trajectory provides a grounded path to future milestones, such as Bell Pair generation, many-qubit control, and scaling to 10s of PIC channels, limited by the size of commercial fiber arrays.

\section{Discussion}
This work covers the multi-layered development of a hardware-digital twin technology stack that comprises a quantum repeater node. We construct critical hardware improvements for future quantum repeaters: a PIC with multiple channels coupled via commercially available fiber arrays, with DC strain tuning achieving up to $99.96\%$ connectivity in future iterations and AC capabilities providing a record SnV$^-$ spin driving rate per square root power of $1.04$ MHz/$\sqrt{\text{mW}}$. Our architecture solves spin control limitations preventing entanglement in previous PIC and CMOS approaches \cite{wan2020large, li2024heterogeneous}, and provides an alternative on-chip spectral tuning technique to arrays of frequency modulators off-chip matching emitted photons from inhomogeneous sources \cite{knaut2024entanglement}.
This is a marked improvement on existing quantum repeater hardware \textit{independent} of yet \textit{complementary} to quantum memory control and entanglement, which may enable higher metropolitan entanglement rates and local entanglement purification for Bell states of increased purity \cite{briegel2000entanglement}.

In addition to novel hardware, we employ a novel methodology--the MPhDT-EM framework--that converts an intractable inverse problem (extracting global emitter parameters from a distribution of emitters with unknown positions) into a tractable inference problem, provided a sufficiently accurate physical simulation of the device strain environment is available. The MPhDT, EM algorithm, and experimental w-PLE dataset supply the simulation, inference engine, and constraints, respectively. The extracted parameters $t_{\perp}$ and $t_{||}$ are then physically meaningful material definitions that transfer across devices and can be used to design future QMCs from first principles. The simultaneous engineering of QR-PIC hardware, MPhDT, and EM- and regression- based inference represents a leap towards full-stack development methodologies for quantum repeaters \cite{awschalom2025challenges, zhou2025opportunities, riedel2026scalable}.

Several aspects of the approach deserve further development. The current EM dataset comprises 189 $\Delta ZPL$ measurements across 30 emitters, which is sufficient for convergence of $t_{\perp}$ and $t_{||}$ as shown by the Monte Carlo analysis but insufficient to train complex per-emitter machine learning models. Expanded datasets from larger QMC implantation runs will enable more sophisticated emitter parameter prediction and progressively tighter uncertainty bounds on extracted parameters. 
In this way, the MPhDT-EM framework is self-referentially improving: as more samples are constructed and tested under the workflow of this work, the emitter dataset enlarges, increasing confidence in the MPhDT performance and improving the predictive power of the MPhDT-regression framework.
Furthermore, the nuclear spin control in this work is limited to polarization of a strongly coupled nuclear spin and detection of weakly coupled spins; future works can demonstrate coherent operation of a per-SnV$^-$ $^{13}\text{C}$ nuclear spin memory bank. Finally, while the current blanket distribution of SnV$^-$ centers in the QMC serves a useful metrological purpose of varying strain distributions by waveguide location, future QMCs fabricated with masked implantation can house emitters concentrated about the waveguide centers, which would sit near the PIC microwave line and permit spin driving on many emitters in the QMC. This, in tandem with the ability to spectrally overlap many emitters via strain tuning, will open the door to on-chip generation of cluster states of SnV$^-$ electron and nuclear spins.

Both the hardware present in this work and the MPhDT methodology are not specific to SnV$^-$ centers nor to diamond. Any solid-state emitter system in which the relevant emitter parameters are incompletely known, and in which a physically accurate simulation of the device environment may be constructed, is a candidate for this approach. This includes group-IV vacancy centers aside from SnV$^-$ (GeV, PbV), as well as quantum dot systems where strain from heteroepitaxial layers or bonded interfaces produces similar parameter uncertainties \cite{shaji2026scalable}.

\section{Acknowledgements}

DISTRIBUTION STATEMENT A. Approved for public release. Distribution is unlimited. This material is based upon work supported by the Under Secretary of War for Research and Engineering under Air Force Contract No. FA8702-15-D-0001 or FA8702-25-D-B002. Any opinions, findings, conclusions or recommendations expressed in this material are those of the author(s) and do not necessarily reflect the views of the Under Secretary of War for Research and Engineering.

The authors wish to acknowledge fruitful discussions with Chao Li, Di Liu, Isaac Harris, Sofia Patomaki, Chaohan Cui, Prajit Dhara, and Pratyush Anand.


\section{Competing interests}
The authors declare no competing interests.

\section{Data and Materials availability}
All of the data that support the findings of this study are reported in the main text and Supplementary Materials.  Source data are available from the corresponding authors on reasonable request.


\printbibliography[heading=none]
\end{refsection}
\begin{refsection}
\section{Methods}
\subsection{PIC Sample Preparation}
The silicon nitride-on-insulator photonic integrated circuits used in this work were produced at MIT Lincoln Laboratories. PIC samples were stealth diced and edge-polished prior to heterogeneous pick and stamping of QMCs. QMCs are fabricated using electron beam lithography of electronic grade diamond from Element Six implanted with tin ions at 350 keV with $10^{11}/\text{cm}^2$ fluence and annealed at 1200 C and $10^{-7}$ mbar \cite{chen2023protocols, chen2024scalable}.

We achieve a high-yield heterogeneous pick-and-stamp procedure for QMC integration following the steps below.
\begin{enumerate}
    \item Break tethers between nanofabricated QMC and parent EG diamond chip using tungsten microprobes on a 3-axis piezomicromanipulator and rotation stage \cite{mouradian2015scalable}.
    \item Pick the QMC using the tungsten microprobe and place onto a 50 $\mu\text{m}$ x 50 $\mu\text{m}$ X-Celeprint micro-transfer printing PDMS stamp pedestal \cite{6985454}.
    \item Transfer the PDMS stamp to a glass slide and flip-mount the slide on a 3-axis sub-micron flexure stage.
    \item Stamp the QMC onto the PIC socket via placement and shearing of the PDMS stamp from the PIC surface.
\end{enumerate}

We mount the PICs with QMCs on custom copper mounts with printed circuit boards for RF and DC wiring into a Bluefors LD250 cryostat.

\subsection{Expectation-Maximization (EM) Algorithm for Determining Global and Latent SnV$^-$ Variables}\label{sm:em_algorithm}

Determining the global strain susceptibilities $t_{\perp}$ and $t_{||}$ poses a challenging problem. If one were to solely analyze the strain tuning of a single emitter,  large variances in the fitted strain susceptibilities would remain due to the inability to localize a single emitter in waveguide cross-sectional space (ion implantation analysis can narrow the distribution in z, but not in y). A smart experimental design with narrow aperture implantation can reduce this uncertainty, but this does not guarantee a perfect strain estimate due to residual position uncertainty. Hence, we apply an EM algorithm to analyze the strain susceptibility over a distribution of $N \sim 30$ emitters under a number $\geq1$ voltage configurations. An initial filter of emitters with ZPL peaks below a brightness threshold (1000 counts, typical noise floor $\sim200-500$ counts) and/or a signal-to-noise ratio threshold (1.0) trimmed an initial yield of 531 emitter keys to 368 emitters for EM fitting. Following this filter, we apply a filter on the emitter PLE traces versus applied voltage to remove ill-fitted ZPL peaks and noisy residuals from the data pool for the EM algorithm. Ultimately, we fed a total of 189 $\Delta ZPL_i$ data points to the EM algorithm across 30 emitters. Each EM iteration uses 90\% of the data for about 170 data points per run. We attribute the low number of emitter data points in the post-PnS w-PLE data (compared to that of the pre-PnS data) to a combination of factors, including noisy scatter off of metal surfaces in the PIC as opposed to the parent diamond when focusing on a QMC; a larger inhomogeneous distribution in the post-PnS data due to strain-based ZPL shifts, which led to many emitter frequencies falling outside of narrow scan ranges in the post-PnS w-PLE for each sample; and other processing variables that may affect SnV$^-$ brightness in the post-PnS data such as chip cleanliness, adverse effects due to FIB cutting of QMC waveguides, etc.

We set up the initial conditions of the EM algorithm by first extracting the strain maps for each emitter and then instantiating initial guesses for each emitter's position and orientation. The data used to perform the EM algorithm consists of the following items for each SnV$^-$ emitter:
\begin{itemize}
    \item An array of float values that give the potential ZPL shifts for the emitter. Each value is extracted by a peak finding algorithm that identifies peaks in the post-PnS PLE spectrum at the site of the emitter in the pre-PnS PLE data. For example, if an emitter were located at $-5$ um relative to the channel 2 waveguide origin, then in the post-PnS data we integrate a 0.5 um region around channel 2, $x_{wg} = -5 um$, following \SMsec{sm:widefield} and perform the peak finding algorithm to find $m$ peaks. This array of values gives the possible ZPL shifts for the emitter and acommodates the possibility of multiple emitters within a roughly diffraction-limited spot in the w-PLE measurement.
    
    \item A set of 4 x 2 two-dimensional arrays of size $N_y \times N_z$, where $N_y$ and $N_z$ are the number of resolved points in the y-axis and z-axis of the waveguide cross section, at the emitter location. The first set of 4 arrays is the $\varepsilon_{zz}(o,y,z)$ array giving axial strain for each of four non-degenerate orientations $o$ (Transverse 1/2, Axial 1/2). The second set of 4 arrays is the $\varepsilon_{xx}(o,y,z) + \varepsilon_{yy}(o,y,z)$ array giving transverse strain for each $o$. This gives the modeled strain shift experienced by the emitter dependent on its orientation and waveguide cross-sectional position.
\end{itemize}

We seed the initial guess of emitter probability position using the Transport of Ions in Matter (TRIM) Monte Carlo software. 
We multiply this by a Gaussian distribution in $y$ with large $\sigma = 140\text{ nm}$ so as to mitigate the EM algorithm fitting non-physical edge positions for each emitter, preferring a location nearer to the center of the cross-section in $y$. This is consistent with the assumption that emitter properties would heavily degrade at dielectric boundaries \cite{martin2024topological}. The output is a 4 x 2D grid of probability distributions over orientation, y, and z.

For simplicity's sake in this work, we consider only two of the four orientations--one transverse orientation and one axial orientation--when performing the EM optimization. This nets us sufficient information regarding the emitters in the system and their tunability under thermo-electromechanical strain. 

We apply a small corrective probability (1e-6) to all points in the probability distribution and renormalize to prevent zero-probability pixels, which would cause errors in the EM algorithm.

With initial conditions in hand, we initiate the algorithm. As stated in the main text, the two algorithm steps are as follows.
\begin{enumerate}
\item{
\textbf{E-step:} Update responsibility weight of each $(o,y,z)$ assignment for emitter $i$ based on agreement with the measured $\Delta ZPL_i$. 
\begin{multline}
        r_{i,o,y,z,t} = \\ r_{i,o,y,z,t-1}\left(\frac{\exp\left(-(V_{i} - \mu_{i}(o,y,z)\right)^2/(2\sigma^2))}{\sum_{o',y',z'}\exp\left(-(V_{i} - \mu_{i}(o',y',z')\right)^2/(2\sigma^2))}\right),
    \end{multline}
where $V_i$ is the measured $\Delta ZPL_{i}$ and $\sigma$ is the noise standard deviation, which we fix at 10 GHz. This assigns higher probability to $(o,y,z)$ configurations that better predict the observed ZPL shift.
}
\item{
\textbf{M-step:} Update $(t_{\perp}, t_{||})$ by minimizing the responsibility-weighted squared residual. 
\begin{equation}
        \Theta_t = \sum_{i,o,y,z}r_{i,o,y,z,t}\left(V_i - \mu_i(o,y,z)\right)^2
    \end{equation}
with respect to $t_{\perp}$ and $t_{||}$. This drives the global parameters toward values that best explain the entire ensemble under the current emitter position estimates.
}
\end{enumerate}

In essence, for each iteration, the E-step modifies the probability distribution of emitters and the M-step modifies $t_{\perp}, t_{||}$ according to the above equations. During the E-step, the probability distribution of the ZPL shift array is also modified, such that the algorithm may converge upon a single value as the likely ZPL shift of the emitter (in the event of multiple peaks detected in the same spatial location). The E-step and M-step are alternated until convergence, typically under 1000 iterations.
\printbibliography[heading=none]
\end{refsection}
\widetext
\newpage
\clearpage
\begin{figure*}
    \centering
    \includegraphics[width=\linewidth]{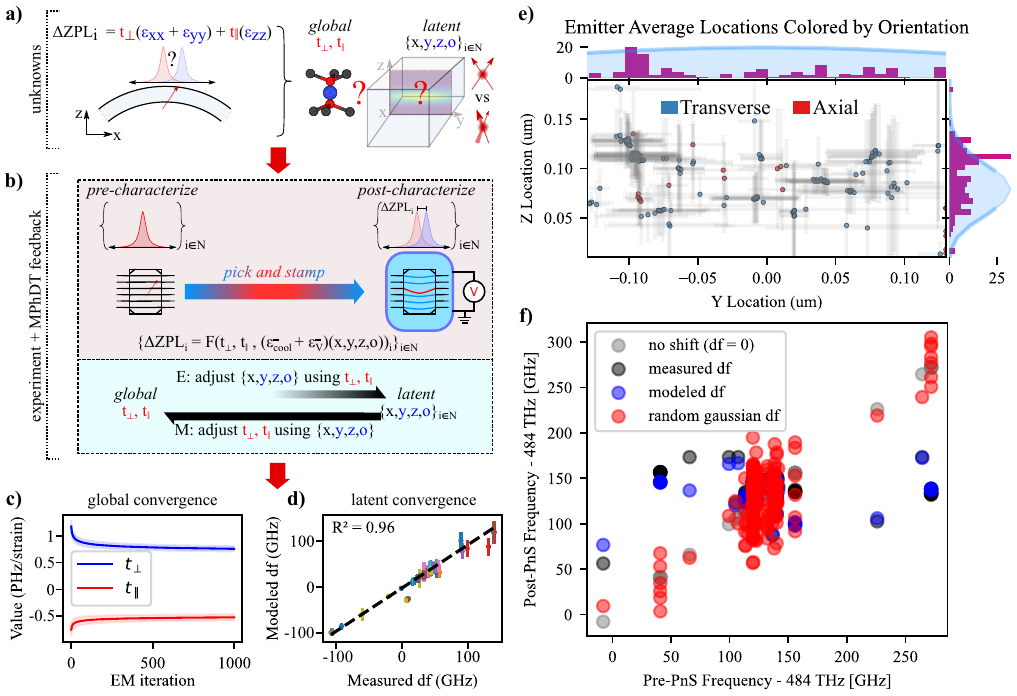}
    \caption{Extended EM maximization figure. (a) illustration of on-chip strain in a diamond waveguide generating frequency shifts in the ZPL of a SnV$^-$ post-PnS and upon applied MEMS voltage. The ZPL shift is difficult to predict \textit{a priori} due to the globally uncertain and underreported variables of $t_{perp}$ and $t_{par}$, as well as the latent variables of each emitter (waveguide cross-sectional location, orientation). (b) Expectation-maximization (EM) algorithm run on seven experimental QMCs of the PnS procedure and applied MEMS voltage on test chips. The ZPL shift of emitters in each sample are observed and recorded, and the collective data is inserted in a MPhDT-informed algorithm that takes simulated cross-sectional waveguide strain at each emitter location with the experimentally observed ZPL shifts and iteratively re-tunes the global variables \{$t_{perp}$, $t_{par}$\} and re-fits the expected \{(y,z,o)\}$_{i \in N}$ of each emitter $i$. (c-d) convergence of the EM algorithm on both global (c) and latent (d) fronts across 100 Monte Carlo algorithm runs. (e) resultant emitter positional distribution in the waveguide cross-section, with histograms in Y and Z overlaid on the assumed distributions. (f) scatter plot of emitter pre-PnS frequencies against post-PnS frequencies, showing a $\sim 70\%$ reduction in modeled $\Delta ZPL$ error in the model (as well as in the XGBoost prediction test data, not shown) compared to no assumed frequency shift or a normally distributed frequency shift with a sigma given by the mean $\Delta ZPL$ in the observed w-PLE data.}
    \label{fig:em_maximization_extended}
\end{figure*}
\clearpage
\newpage

\begin{refsection}
    
\section*{Supplementary Materials for Full-Stack High-Volume Quantum Networking Architecture based on Photonic-Integrated Tin Vacancy Centers in Diamond}


\section*{Supplementary Table of Contents}
\vspace{-0.75em}
\hrule
\vspace{0.9em}

\newcommand{\SupTOCEntry}[3]{%
  #1.\quad\hyperref[#2]{#3}%
  \dotfill\pageref{#2}\par
}

\begingroup
\setlength{\parindent}{0pt}
\setlength{\parskip}{0.35em}


\SupTOCEntry{S1}{sm:widefield}
  {Widefield Photoluminescence Excitation and Emitter Analysis}
\SupTOCEntry{S2}{sm:PnS}
  {Heterogeneous Pick-and Stamp for Integrating Quantum Microchiplets}
\SupTOCEntry{S3}{sm:MPhDT}
  {Digital Twin Construction in FEM/FDTD Using OpenCV}
\SupTOCEntry{S4}{sm:em_algorithm}
  {Expectation-Maximization (EM) Algorithm Monte Carlo Sampling}
\SupTOCEntry{S5}{sm:pulseseq}
  {Pulseseq: A Package for RFSoC-Enabled Quantum Control}
\SupTOCEntry{S6}{sm:xgboost}
  {Extreme Gradient Boosting Supervised Learning Applied to SnV$^-$ Emitter Prediction}
\SupTOCEntry{S7}{sm:exp-bf}
  {Experimental Details of Optical and Microwave Control in a Bluefors Cryostat}
\SupTOCEntry{S8}{sm:strain-tuning}
  {Electromechanically Strain-Tuned Photoluminescence Excitation}
\SupTOCEntry{S9}{sm:spin-echo}
  {Nuclear Spin Detection and Spin Memory Quantification via Electron Spin Echo}
\SupTOCEntry{S10}{sm:future-devices}
  {Exponential Scaling Prospects of QR-PICs}




\endgroup



\begin{supplement}

\newpage



\section{Widefield Photoluminescence Excitation and Emitter Analysis}\label{sm:widefield}

SnV$^-$ emitters in our quantum microchiplets are characterized using widefield photoluminescence excitation (PLE) in the setup shown in \cite{sutula2023large}. A resonant laser (Msquared SolSTiS + EMM Module) is used to sweep a frequency range of $\sim50$ GHz about a central frequency of 484.12 THz (corresponding to 619.25 nm) to detect zero-phonon lines about the typical SnV$^-$ ZPL value \cite{trusheim2020transform}. A DC bias voltage is supplied to the birefringent crystal of the EMM, allowing for a temporal frequency chirp over 30 GHz. At time steps separated by $\sim 1$ ms, an EMCCD is used to snap signal images of the QMC under test through a free space collection path under exposure of this chirped resonant laser and constant green (Verdi 532 nm laser) excitation, where the signal is the spatially confined phonon sideband of excited emitters in the QMC (filtered with a green filter and a ZPL filter in the collection path). Both lasers are operated in continuous wave mode; approximately 70 uW of green power is used, while approximately 10 uW of resonant power is used, across scans of 30 microchiplets in a bulk electronic grade parent diamond. The images are saved to a .mat file and compressed into .h5 files for processing in Python.

\subsection{Emitter Characterization from w-PLE}

Because we are interested in SnV$^-$ emitters in the photonic waveguides of the QMC as opposed to emitters in the QMC frame or in bulk diamond, we post-process the collected data to develop "spectrographs" of the QMC (Fig.~\ref{fig:SI_widefield_ple}). These spectrographs are a set of six 2D arrays of counts vs. x-location in the waveguide and frequency, for each of six waveguides in the fabricated QMCs.

\begin{figure}
    \centering
    \includegraphics[width=\linewidth]{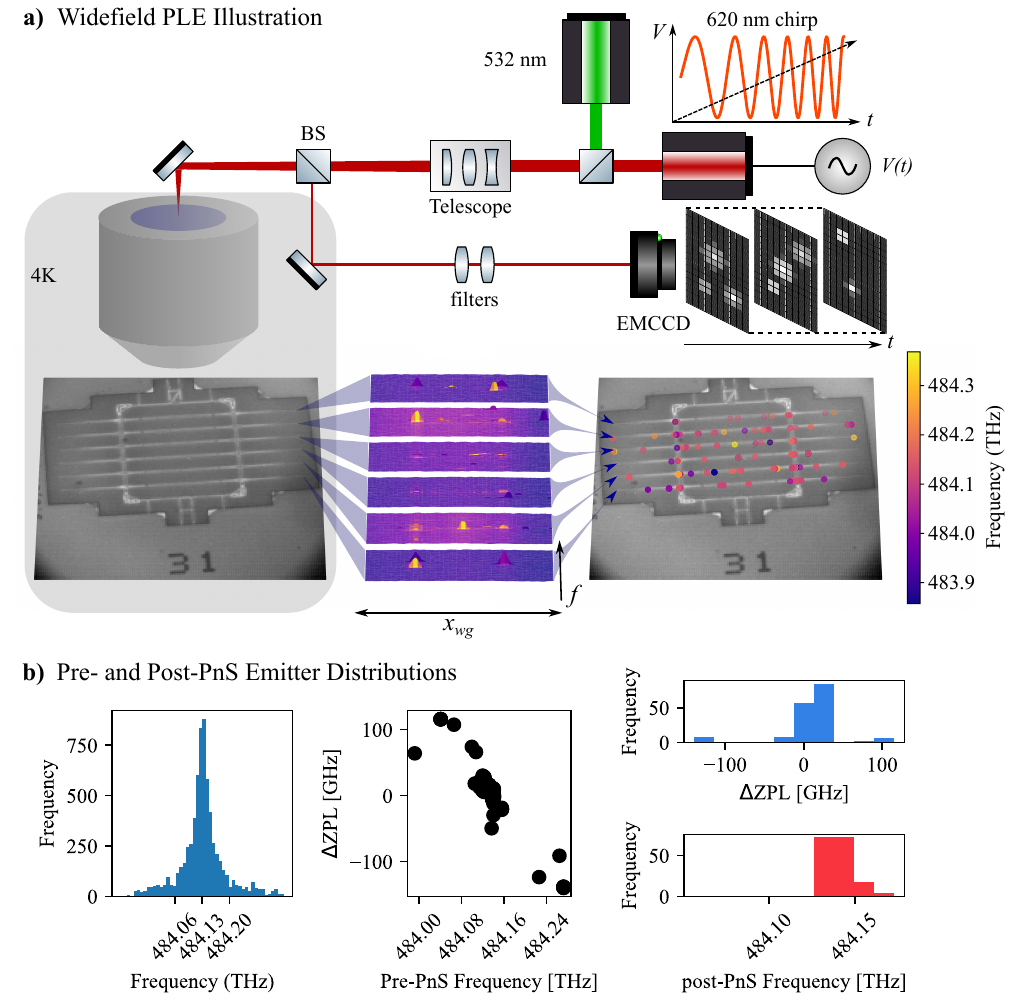}
    \caption{Widefield PLE (w-PLE) of QMCs. (a) illustrates the w-PLE optical schematic and the processing of spectrographs for a QMC, from which we extract emitters via peak-finding. (b) Inhomogeneous distribution of the pre-PnS w-PLE emitter data (left); scatter plot of initial frequency against $\Delta ZPL$ (middle); and distributions of $\Delta ZPL$ and post-PnS frequency of post-PnS data (right).}
    \label{fig:SI_widefield_ple}
\end{figure}

From each QMC spectrograph, we extract emitter spectral and spatial locations by applying the following algorithm:
\begin{enumerate}
    \item Apply a black-white threshold to determine emitter locations
    \item determine the centroids of each emitter location
    \item employ a peak finding algorithm to identify SnV$^-$ signatures in each centroid
    \item Extract the following parameters of the emitter from each peak:
    \begin{itemize}
        \item unique emitter ID $i$;
        \item central frequency $f_{0_i}$;
        \item spatial location $x_{wg_i}$;
        \item linewidth $\gamma_i$;
        \item "central" brightness at the $x_{wg_i}$ location $A_{c_i}$;
        \item left and right "tip" brightness at the spatial edges of the spectrograph, at the central frequency of the emitter $A_{L_i}, A_{R_i}$;
        \item noise floor amplitude $A_{N_i}$;
        \item goodness of fit metrics.
    \end{itemize}
\end{enumerate}

A similar technique is applied in \cite{duan2025bayesian}.

\subsection{Emitter Tracking pre- and post-PnS}

For each pick-and-stamped QMC, the above analysis is conducted on the pre-PnS and post-PnS w-PLE datasets. Then a pre-post analysis is conducted to locate emitters before and after stamping. We catalogue the emitters that are identified in the post-PnS data and matched to a pre-PnS emitter to pass as compiled emitter data to the QR-PIC digital twin. We present the overall pre-PnS ZPL inhomogeneous distribution, as well as the filtered dataset of emitters used in the EM algorithm described in the main text, in Fig.~\ref{fig:SI_widefield_ple}b.

\section{Heterogeneous Pick-and Stamp for Integrating Quantum Microchiplets}\label{sm:PnS}

We heterogeneously integrate QMCs onto the PIC via a pick-and-stamp procedure. First, we use a tungsten probe to break the tethers of the QMC. Three-axis servo and piezo controllers are used to position the tungsten probe at each tether position, and then the piezo knobs are used to push the probe against the tethers to break them. This process is repeated for all tethers at the top and bottom of the QMC (Fig.~\ref{fig:SI_widefield_ple} QMC image--four on top and four on bottom).

Following this process, two possibilities may occur. The first is that the QMC adheres to the tungsten probe to "pick" it from the parent diamond, as in \cite{wan2020large}. The second is that the QMC falls into the quasi-isotropically underetched trench below and is not recoverable by the tungsten probe. In the second case, a PDMS stamp is used to contact the topside of the QMC and "pick" it out of the trench. Due to the recoverable nature of the QMC using a stamp, this "pick" step has near-unity yield across O(10) QMCs, excepting human error. We place the picked QMC onto a 50 um x 50 um polydimethylsiloxane (PDMS) stamp, where the square stamp extends above an amorphous PDMS substrate on a glass slide. This placement also poses near-unity yield excepting human error.

We accomplish the "stamp" process by flipping the glass slide such that the PDMS stamp is on the bottom side (QMC topside facing the ground) and mounting it on a three-axis Thorlabs Nanomax positioner. We then "stamp" the QMC into a PIC socket by pressing it down onto the PIC, aligning the diamond waveguides to SiN waveguide tapers using a Thorlabs camera to view the QMC through the transparent stamp, and shearing the stamp away from the substrate. Importantly, if we find that the placement is not accurate (i.e. the waveguides are misaligned by $\sim$0.5 $\mu\text{m}$ or greater distance), then we can re-pick and re-stamp the QMC with the flipped PDMS stamp. This results in a near-unity yield stamping process. The high yield of the overall integration technique allows us to evaluate the procedure on the "coupling capture" efficiency outlined in the main text as opposed to the QMC loss during the pick-and-stamp procedure.

We note that a similar technique was employed in \cite{harris2025high}, in which the sample was prepared in parallel and collaboration with this work.

\section{Digital Twin Construction in FEM/FDTD Using OpenCV}\label{sm:MPhDT}

In accompaniment to each sample, we construct a multiphysics digital twin (MPhDT). This consists of a collection of files in COMSOL Multiphysics, Tidy3D, and Python that define the unique sample geometry given the PnS position of the QMC and given pre-characterized emitter locations.

Each MPhDT begins from a base file that is loaded with unique parameters from experimental data (pick-and-stamp position, SnV$^-$ waveguide and channel positions pre-characterized under w-PLE). 

\subsection{COMSOL FEM Twin Construction (Thermo- and Electro-mechanics, Electromagnetics)}\label{sm:MPhDT:FEM}
\begin{figure}
    \centering
    \includegraphics[width=\linewidth]{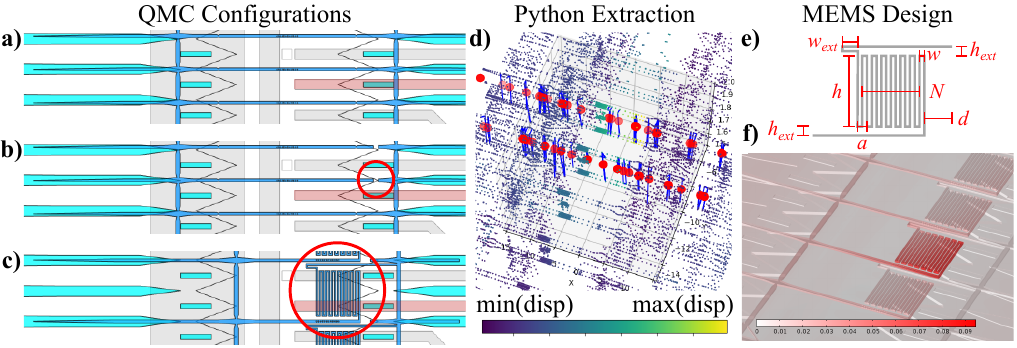}
    \caption{Illustration of electromechanical tuning in COMSOL. (a-c) depict the different QMC configurations simulated in this work on one QR-PIC type featured in this work, from (a) default to (b) FIB cut spotlighted to (c) MEMS structure spotlighted. The red electrode is set to a DC value $0 \leq V_{DC} \leq 100$ and the gray electrodes are grounded. (d) illustrates the data extraction and reconstruction in Python, where each dot (non-red) illustrates a mesh point and its associated mechanical displacement under thermal contraction. The red dots indicate SnV$^-$ locations in the waveguide, and the blue polygons indicate nearest mesh points used to interpolate simulation values at each emitter location. (e) depicts the MEMS design considered in this work and relevant parameters, and the example panel (f) illustrates the mechanical displacement profile in COMSOL for one QMC configuration (MEMS, N=13) under 100V.}
    \label{fig:SI_COMSOL_FEM}
\end{figure}

\begin{figure}
    \centering
    \includegraphics[width=\linewidth]{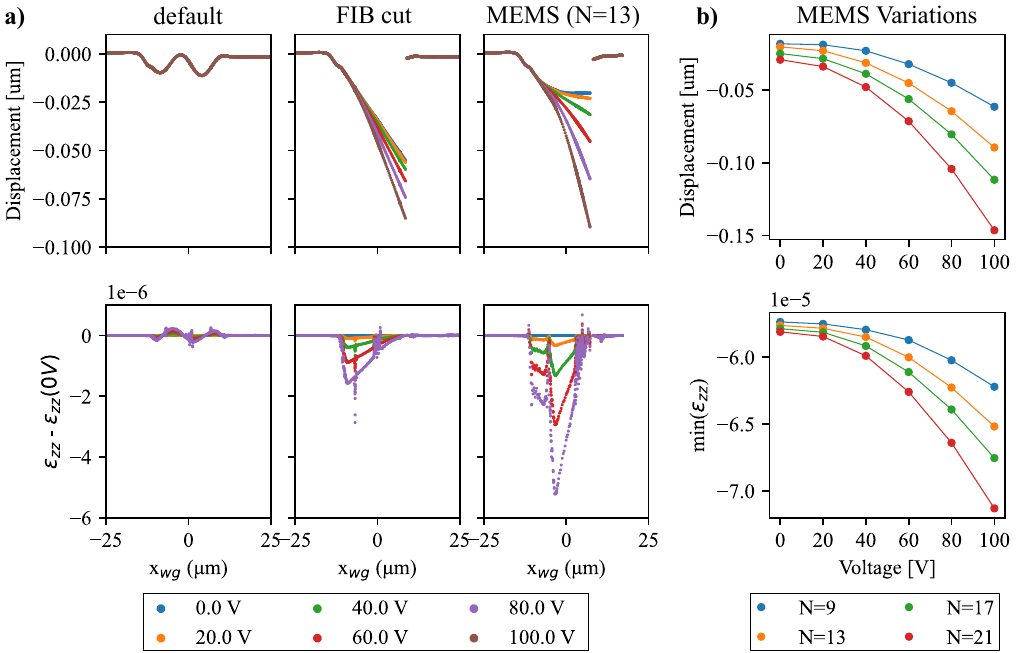}
    \caption{(a) Displacement and strain profiles of three diamond nanobeam waveguide configurations in this work. (b) Further simulations of the MEMS nanobeam configuration under different variations of $N$ cantilever periods.}
    \label{fig:SI_mems_tuning_plots}
\end{figure}

\begin{figure}
    \centering
    \includegraphics[width=\linewidth]{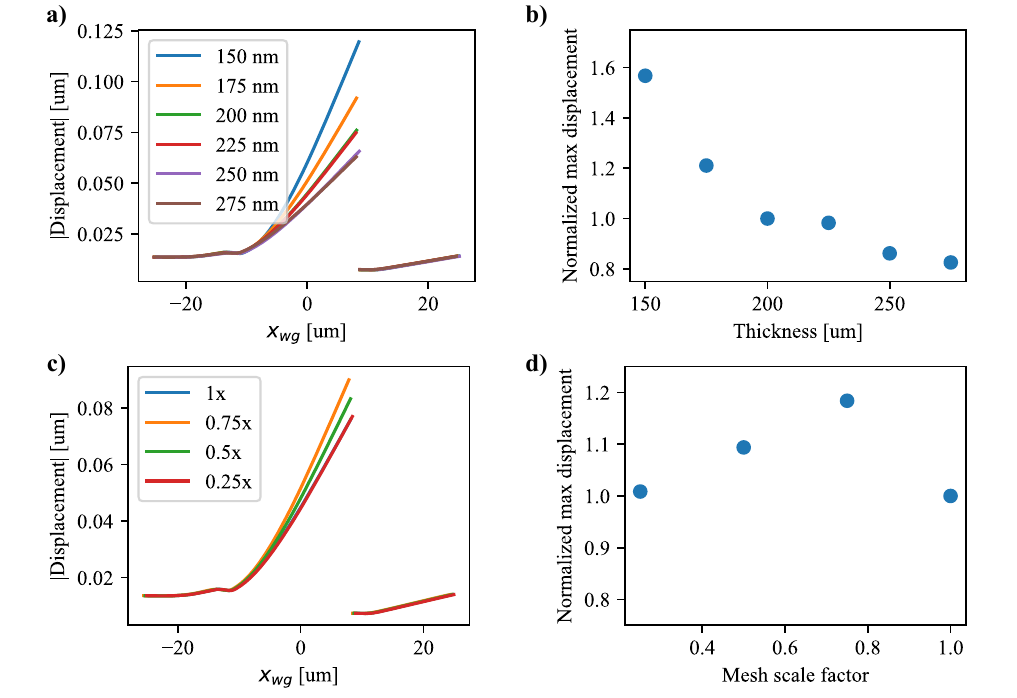}
    \caption{(a) Displacement profile of FIB cut diamond nanobeam waveguides of different thicknesses, with (b) depicting the normalized maximum displacement of the beam with respect to a reference thickness of 200 nm. (c) and (d) depict the displacement profile of the 200 nm beam under COMSOL mesh variations, where the largest mesh element of 8.54 um and smallest mesh element of 1.54 um are scaled by the scale factor. Both simulations show that the COMSOL MPhDT is tolerant to minor variations in diamond thickness and simulated mesh quality.}
    \label{fig:COMSOL_FEM_simulation_variations}
\end{figure}

In COMSOL, the base file consists of a SiN PIC defined from a GDSII file and layer stack imported using COMSOL's AutoCAD import feature and layer specification. In this work, the following modules are implemented:
\begin{itemize}\setlength{\itemsep}{0pt}\setlength{\parskip}{0pt}\setlength{\parsep}{0pt}
    \item Electrostatics (es)
    \item Structural Mechanics (solid)
    \item Heat Transfer in Solids (ht)
    \item Magnetic Fields (mf)
    \item Electromechanical Forces (eme)
    \item Thermal Expansion (te)
\end{itemize}

To simulate chip-scale effects, roughly $(l_x, l_y) = (54, 40)\text{ um}$ of the CAD surrounding the socket origin is defined in the COMSOL geometry, and continuity and periodic boundary conditions are implemented on the left-right and top-bottom edges of the socket, respectively.

The QMC is imported from GDS using the AutoCAD import and its position relative to the PIC socket origin is parametrized with the offsets $(dx, dy, d\theta)_{Q-P}$. We cover the extraction of these parameters in the subsection \SMsec{sm:MPhDT_analysis} below.

We simulate the electromechanical performance of the chip by running COMSOL's Stationary Solver on the sample under varying voltage conditions. We determine the zero-voltage cooldown behavior of the QR-PIC by solving the stationary configuration of the sample at 4 Kelvin given an initial, unstrained condition at 300 Kelvin. This assumes that any strain imparted to the microchiplet during the PnS at 300 K is much less than that imparted by thermal expansion mismatches in the PIC and QMC upon cooldown to 4 K. Electrodes buried in the oxide layer of the QR-PIC about 700 nm below the surface of the QMC provide electromechanical tuning by changing the energy density in the sample, with focus on the QMC waveguide as a diamond nanobeam under electric field, according to the equation
\begin{equation}
    H_{eme} = W_s(\textbf{C}) - \frac{1}{2}\epsilon_o\epsilon_r J \textbf{C}^{-1}:(\textbf{E}\otimes\textbf{E}).
\end{equation}
Here, $H_{eme}$ is COMSOL's energy density as a function of electromechanical effects \cite{COMSOL_StructuralMechanicsModule, COMSOL_Electromechanics_Theory_2026}. The first term, $W_s$, is the mechanical strain energy of the system, while the second term is the electrostatic energy of the system, modified by deformation. The right Cauchy-Green tensor, $\textbf{C} = \textbf{F}^T\textbf{F}$, is a measure of strain in the system ($F$ is the deformation gradient) and $J$ is a measure of the change in volume. Finally, $\textbf{E}$ is the electric field generated by the applied voltage. COMSOL solves this expression for all DC voltage configurations of a selected electrode.

Note that mechanical stress in the system is given by $\textbf{S} = 2\frac{\partial H_{eme}}{\partial \textbf{C}}$, providing us the intuition that, for a linear elastic system approximation, the QMC waveguide strain scales quadratically with electric field and therefore applied voltage $V_{app}$. We therefore explore the quadratic strain tuning curves of our sample in \SMsec{sm:strain-tuning}.

\subsection{Tidy3D Twin Construction (Optics/Photonics)}\label{sm:FDTD}

\begin{figure}
    \centering
    \includegraphics[width=\linewidth]{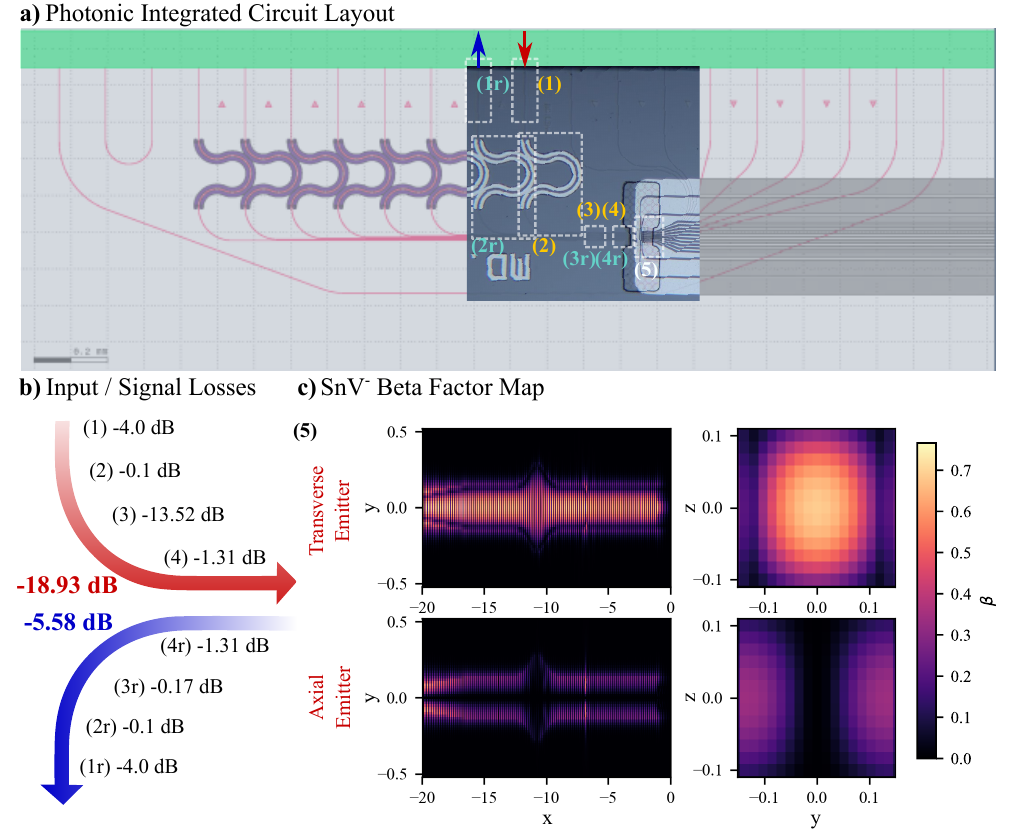}
    \caption{Summary of the QR-PIC photonics deployed in the Bluefors cryostat. (a) PIC layout with optical microscope image of a sample PIC overlayed on the CAD geometry. (1/1r) indicate edge couplers designed for coupling to an edge-coupled, single mode fiber array; (2/2r) indicate TE polarizers; (3/3r) indicate directional couplers (97:3 at 620 nm); (4/4r) indicate the SiN-to-QMC waveguide couplers described in the main text Fig.~\ref{fig:sample_construction}; and (5) indicates the QMC location with implanted SnV$^-$ vacancies. (b) estimated losses at each photonic component. (1) includes alignment errors between the fiber array and PIC edge. (4) estimates the SiN-to-QMC waveguide coupling but does not account for losses accumulated at the interface of the SiN waveguide with an oxide etch that permits heterogeneous QMC integration. (c) Simulated beta factors ($\beta$) along the QMC waveguide (left column) and in a cross-section at a beta factor anti-node along the waveguide for transverse and axial emitter orientations. The sinusoidal $\beta$ along the waveguide is caused by a Bragg reflector in the center of the QMC in deployed samples, resulting in destructive interference for periodic dipole locations along the waveguide.}
    \label{fig:SI_photonics}
\end{figure}

The interference pattern of the cavity reflection creates periodic constructive and destructive interference of the emitter with itself, leading to a nonuniform $\beta$ factor inside of the straight waveguide section. We model this by exciting a standing wave inside of the waveguide. We then simulate the $\beta$ factor and Purcell enhancement $F_p$ of an optimally aligned dipole placed in the strongest field location in the mode using FDTD. From this, we extract a factor $F_r$ relating to the dipole loss into free space modes as~\cite{martin2024topological}
\begin{equation}
    F_r = F_p\frac{(1-\beta)}{\beta}.
\end{equation}
We calculate the Purcell enhancement at point $r$ as 
\begin{equation}
    F_p(r, \hat{n}) = F_{p,max}\frac{|E(r)\cdot \hat{n}|^2}{|E(r_{max})|^2},
\end{equation}
and finally, the local beta factor,
\begin{equation}
    \beta(r,\hat{n})=\frac{F_p(r,\hat{n})}{F_p(r,\hat{n})+F_r}.
\end{equation}
We plot the beta factor for the transverse and axial emitters as a function of space in Fig.~\ref{fig:SI_photonics}c.

\subsection{Sample-Dependent Construction and Analysis using Python OpenCV}\label{sm:MPhDT_analysis}
We accomplish sample-dependent analysis using a series of Python .py files and .ipynb notebooks that access the pre-defined COMSOL FEM and Tidy3D FDTD modules in headless simulations. First, we use Python's OpenCV package to load experimental w-PLE data for a sample. The values $(dx, dy, d\theta)_{Q-P}$ are extracted using Python's OpenCV package from a white-light image capture taken before each sample's w-PLE pre-characterization is initiated. The PIC metal layer template and the QMC template are used in a template match with pixel-resolution in $x,y$ and 0.5 degree resolution in $\theta$ to determine the QMC $(x,y,\theta)_{Q}$ and PIC $(x,y,\theta)_{P}$ with respect to the image origin. The MPhDT parameters $(dx, dy, d\theta)_{Q-P} = (x_Q-x_P, y_Q-y_P, \theta_Q-\theta_P)$. The diffraction limited resolution is given by the Rayleigh criterion for a white light source going through a microscope objective with numerical aperture of 0.9, taken to be approximately $r = \frac{0.61\lambda}{NA} \sim \frac{0.61(550\text{ nm})}{0.9}=373\text{ nm}$. For a $(\sim50\;\mu\text{m})^2$ image region, this gives an expected localization precision of $\sim C\cdot373\text{ nm}/\sqrt{N}$, where $N = (50\;\mu\text{m}/373 \text{ nm})^2\approx134$ is the number of independent resolution elements across the template and $C\sim10-20$ is a correction for cross-correlation sub-pixel accuracy \cite{thompson2002precise,mortensen2010optimized,guizar2008efficient}. This resolution is sufficient to resolve the microchiplet within the high ($>0.5$) power transmission band of the contour plot in Fig.2d of the main text, i.e. whether the QMC waveguide sufficiently overlaps with the SiN channel's waveguide taper. The $(dx, dy, d\theta)_{Q-P}$ across all samples, as well as pre- and post-PnS w-PLE-characterized SnV$^-$ emitters, are saved to a local database for efficient access and sample construction. Fig.~\ref{fig:SI_template_matching} shows a visualization of the template matching results.
\begin{figure}
    \centering
    \includegraphics[width=0.9\linewidth]{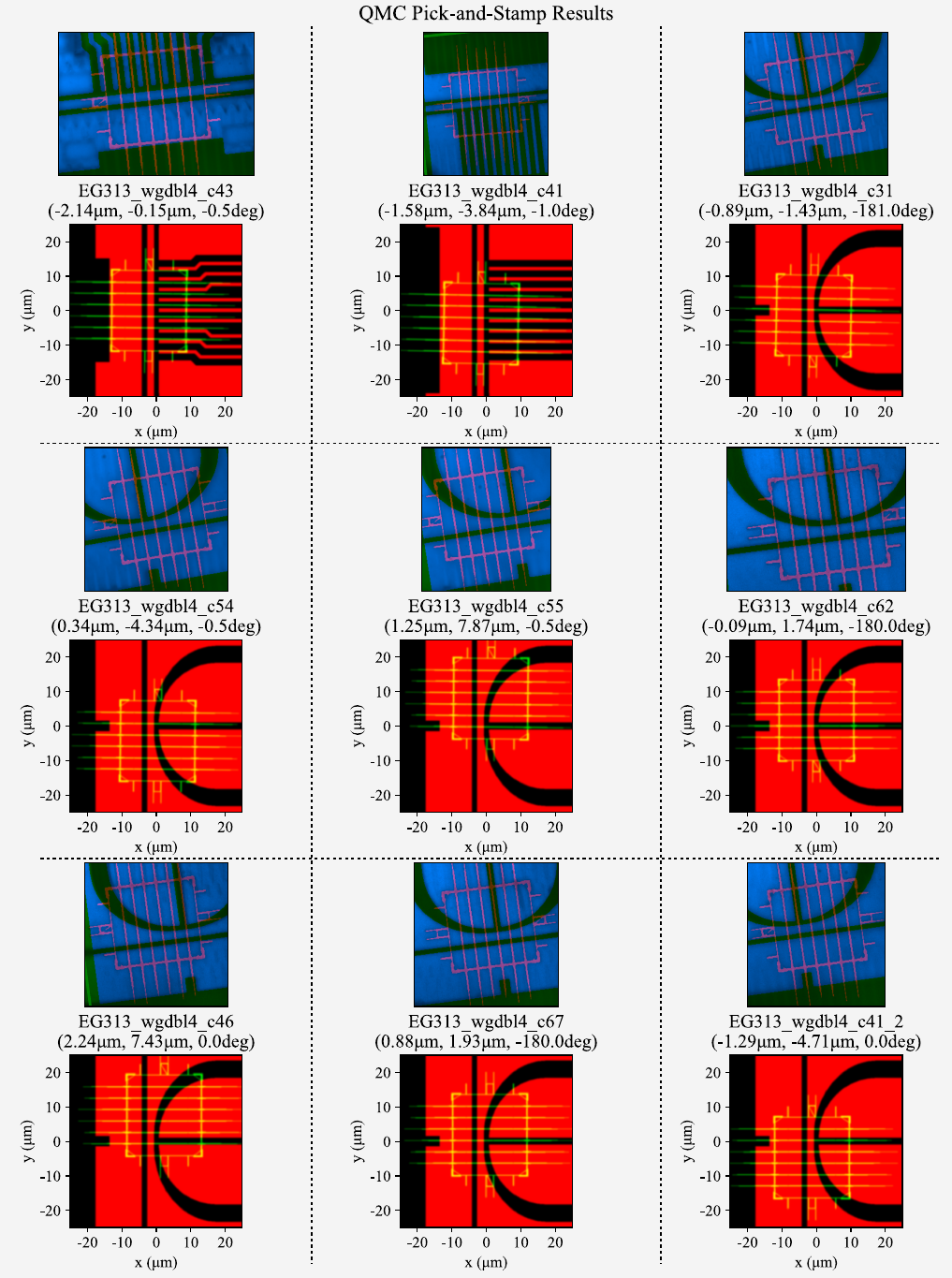}
    \caption{Results of OpenCV template matching for example samples in this work. The top figure of each sample is an RGB microscope image (R=QMC template overlay, G=white light, B=PIC metal template overlay), and the bottom figure in each sample is an RGB image of the template results passed to FEM (R=metal template position centered at the origin, G=QMC offset position). The labels indicate the sample ID and $(dx, dy, d\theta)$ values.}
    \label{fig:SI_template_matching}
\end{figure}

Once we have loaded a headless simulation with the individual sample parameters, we extract goal-dependent data from the MPhDT. When evaluating photonics-scale performance, we are primarily concerned with the power transmission of a mode excited in a QMC waveguide to a mode collected in the coupled SiN channel. We excite the fundamental TE-like mode of the diamond waveguide and monitor the fundamental mode of the SiN. Initially, we extract simulated data by calculating the power transmission as a function of QMC $dy, d\theta$ for fixed $dx=0$, generating the contour plot in the main text. Then, we calculate the power transmission per waveguide per sample and plot these points as markers on the contour plot in the main text. We note that, due to a pitch mismatch of about 280 nm between the QMC waveguides and PIC waveguides resulting from sample constraints, only a fraction of the waveguides of a QMC are coupled to the PIC at any given time. In fact, this benefits the full-stack development of the QR-PIC architecture by providing us with a variety of conditions for the QMC waveguides studied under widefield, such as (1) double-sided contact with the SiN tapers of the socket, (2) single-sided contact, or (3) free-floating placement due to complete pitch mismatch, generating unique strain environments for the emitters studied in post-PnS w-PLE.

Moving on to the emitter level, we primarily analyze two metrics: (1) strain at each SnV$^-$ emitter's waveguide location and (2) estimated SnV$^-$-to-QMC waveguide beta factor. To evaluate (1), we extract the tetragonal mesh from the electromechanical stationary solution as well as the strain tensor $\overline{\overline{\varepsilon}}(x,y,z)$ at each location. For a given SnV$^-$, we generate a 2D map of points $x_i = x_{wg,i}, y_{ij} \in \{-w_{wg}/2, w_{wg}/2\}, z_{ij} \in \{-t_{wg}/2, t_{wg}/2\}$ perpendicular to the QMC waveguide axis. Here, $x_{wg,i}$ is the emitter location along the QMC waveguide, $w_{wg} = 280\text{ nm}$ is the waveguide width, and $t_{wg} = 200\text{ nm}$ is the waveguide thickness. At each point, we interpolate the strain tensor $\overline{\overline{\varepsilon}}(x,y,z)_{ij}$ using Delaunay triangulation. Thus, we extract the cross-sectional strain map for the EM algorithm outlined in Methods.

 \section{Expectation-Maximization (EM) Algorithm Monte Carlo Sampling}\label{sm:em_algorithm}
\begin{figure}
    \centering
    \includegraphics[width=\linewidth]{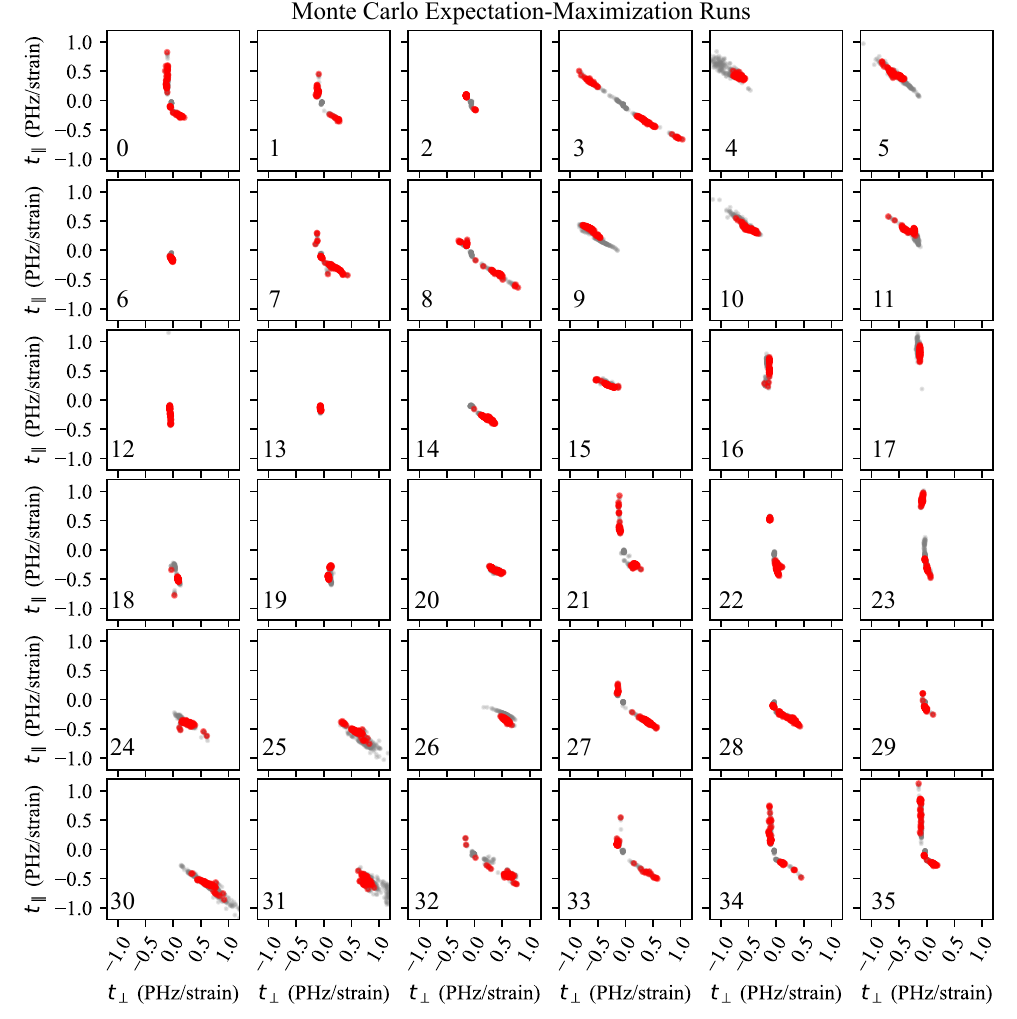}
    \caption{Progression of $t_{\perp}, t_{||}$ over the course of the EM algorithm optimization for each grid point. Red points indicate the end result of each of 100 trials, and gray points indicate the progression to the endpoint every 100 iterations. The Monte Carlo run number is listed in the top left corner of each scatter plot.}
    \label{fig:EM_results_per_MC_run}
\end{figure}

\begin{figure}
    \centering
    \includegraphics[width=\linewidth]{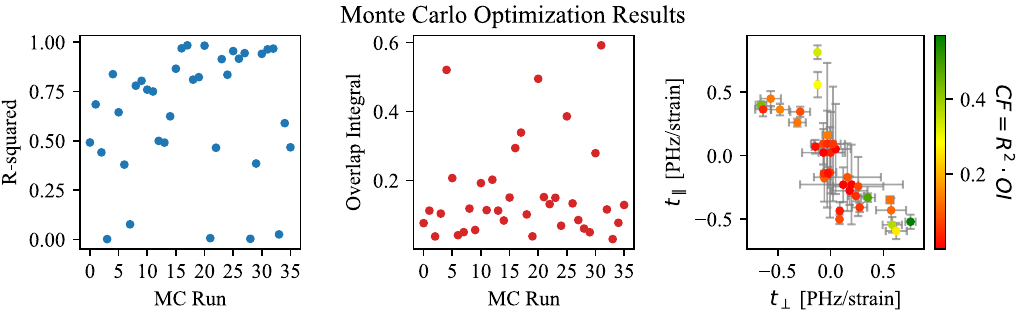}
    \caption{Monte Carlo optimization results by iteration against the evaluation metrics of R-squared value, overlap integral with the TRIM expected vertical emitter distribution in diamond, and cost function $CF = R^2\cdot OI$.}
    \label{fig:EM_results_cost_function_plotting}
\end{figure}

Here, we cover the Monte Carlo sampling variations in the EM algorithm outlined in Methods. To identify the SnV$^-$ strain susceptibilities in a 2D space with many potential local maxima, we perform the EM algorithm optimization in 36 independent runs with initial conditions of $t_{\perp}$ and $t_{||}$ selected from a 6 x 6 array, where $t_{\perp}\in[-2,2]$ PHz/strain and $t_{||}\in[-2,2]$ PHz/strain at even increments. In each optimization run, we perform 100 trials which progress for 1000 iterations as outlined in Methods; these trial results are averaged for the per-run Monte Carlo result.

In each trial, we apply the following variations on the strain map and initial conditions, found in the EM optimization code (see Data Availability):
\begin{itemize}
    \item STRAIN\_PERTURBATION\_FRACTION = 0.20: Gaussian multiplicative perturbation on each strain map pixel, accounting for the $\sim20\%$ uncertainty in COMSOL simulations due to mesh non-ideality (Fig.~\ref{fig:COMSOL_FEM_simulation_variations}) \cite{song2009effect, buzzi2025spectral}.
    \item STRAIN\_ADDITIVE\_PERTURBATION\_FRACTION = 0.10: Gaussian additive perturbation on each strain map pixel, accounting for potential variations in diamond thickness, or inconsistencies with the physical experiment.
    \item INITIAL\_TPERP\_TPAR\_PERTURBATION\_FRACTION = 0.10: Gaussian multiplicative perturbation on initial $t_{\perp}, t_{||}$ conditions for each per-run trial.
    \item INITIAL\_TPERP\_TPAR\_ADDITIVE\_PERTURBATION = 0.02e15: Gaussian additive perturbation on initial $t_{\perp}, t_{||}$ conditions for each per-run trial.
    \item EMITTER\_SAMPLE\_FRACTION = 0.90: fraction of emitters used in each EM optimization trial.
\end{itemize}

Fig.~\ref{fig:EM_results_per_MC_run} shows the results of the EM algorithm for each run, identifying local minima near the origin where the EM algorithm heavily biases emitter locations to the edges of the diamond nanobeam instead of tuning the strain susceptibilities to appropriate values. In selecting the optimal EM result, we determine the winning run by evaluating each Monte Carlo result against a cost function
\begin{equation}
    CF = R^2\cdot OI,\; OI = \frac{\left( \int_{z_{min}}^{z_{max}} \tilde{f}_{\mathrm{measured}}(z) \tilde{f}_{\mathrm{expected}}(z)\, dz \right)^2}{\sqrt{\left(\int_{z_{min}}^{z_{max}} \tilde{f}_{\mathrm{measured}}(z)dz\right)^2\left(\int_{z_{min}}^{z_{max}} \tilde{f}_{\mathrm{expected}}(z)dz\right)^2}},
\end{equation}
where $r^2$ is the r-squared fit between measured and modeled $\Delta ZPL_i$ values, $\tilde{f}_{\mathrm{measured}}(z)$ is the emitter distribution in diamond vertical direction $z$ resulting from the EM optimization, and $\tilde{f}_{\mathrm{expected}}(z)$ is the TRIM-predicted distribution in $z$, which is taken as a ground truth.

\section{Pulseseq: A Package for RFSoC-Enabled Quantum Control}\label{sm:pulseseq}

We performed all waveform generation and measurements using a single Xilinx ZCU111 RFSoC, which we program with a state-machine style software package named Pulseseq. Experiments are built using a hierarchy of classes that enable modular programming, as shown in Figure~\ref{fig:pulseseq}, and can execute decisions in real time based on photon counts received during a state in the state machine \cite{humphreys2018deterministic}.

\begin{figure*}
    \centering
    \includegraphics[width=\textwidth]{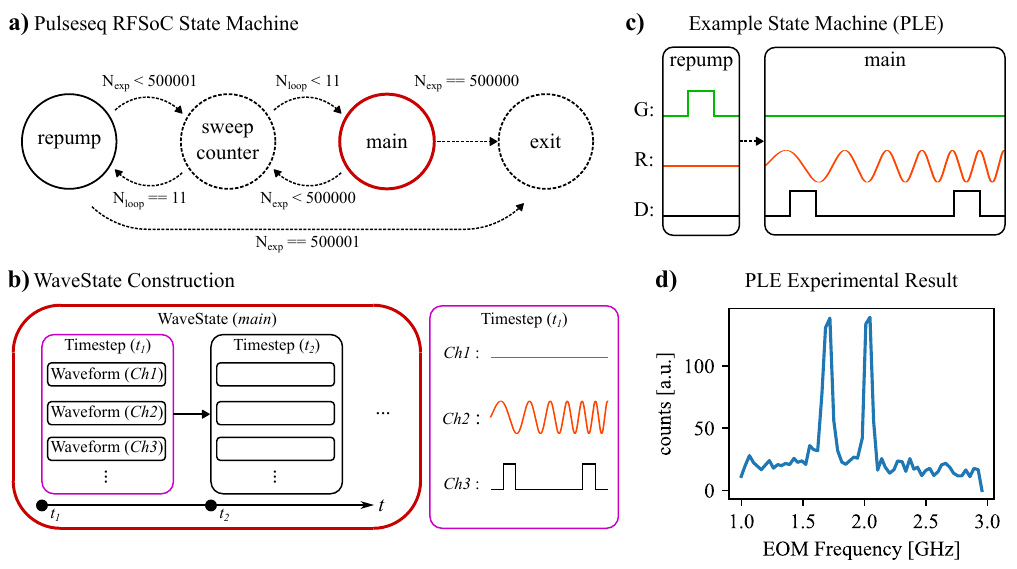}
    \caption{(a) A state machine diagram for chirped PLE. (b) A conceptual diagram of the workflow for playing waveforms in a state of the state machine. Repump only occurs after 11 successive chirps, controlled by the sweep\_counter state. (c) A minimal plot of waveforms that might play in a chirped PLE experiment. (d) Result of a chirped PLE experiment on a SnV center under Zeeman splitting.}
    \label{fig:pulseseq}
\end{figure*}

The Waveform class contains metadata pertaining to generating an arbitrary Waveform. We assign multiple simultaneous Waveform objects a channel in order to create a Timestep object. The WaveState class instructs multiple Timestep objects to play back-to-back, and programs the RFSoC to track the photon counts received during the associated time period. Finally, the WaveMachine class assigns a starting WaveState, and programs the RFSoC to transition between WaveStates based on repetition or photon count requirements.

This modular package allows us to easily program readout-dependent logic such as a charge check into an experiment.

\section{Extreme Gradient Boosting Supervised Learning Applied to SnV$^-$ Emitter Prediction}\label{sm:xgboost}

\begin{figure}
    \centering
    \includegraphics[width=\linewidth]{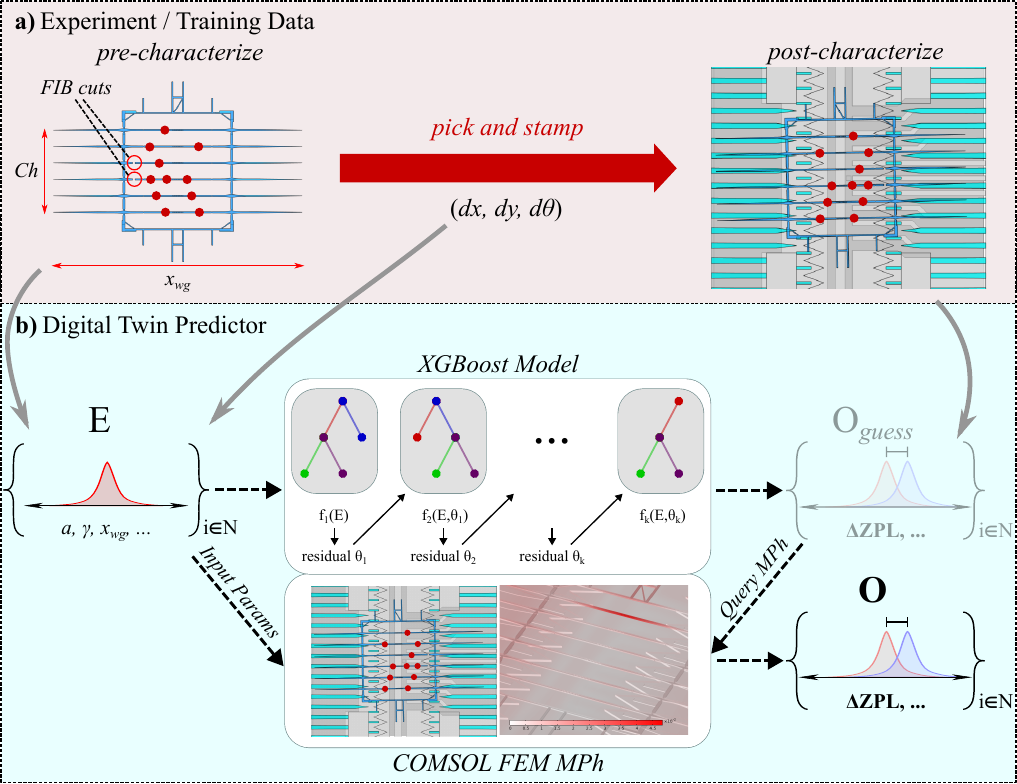}
    \caption{ML Predictor workflow. (a) Gray arrows from the w-PLE data indicate input and output variables used to train an XGBoost regression model. (b) MPhDT-assisted $\Delta ZPL_i$ predictions are achieved by predicting a $\Delta ZPL_i$ via a regression model, and then fine-tuning that guess using a FEM simulation of the cross sectional strain in the waveguide for a given emitter.}
    \label{fig:placeholder}
\end{figure}

To demonstrate the principle of applying a machine learning algorithm on the sample construction pipeline, we employ the Extreme Gradient Boosting (XGBoost) algorithm for SnV$^-$ ZPL shift prediction after PnS. The XGBoost algorithm is a class of gradient boosting machine algorithms that performs regression and/or classification using an ensemble of trees, where each subsequent tree fine-tunes the prediction of the previous one. A detailed treatment of XGBoost can be found \hyperlink{https://xgboost.readthedocs.io/en/stable/tutorials/model.html}{here}.

In our model, the input variables are the initial frequency $f_{ZPL,0_i}$, the emitter brightness $a_i$, and the initial linewidth $\gamma_{i}$ of a SnV$^-$. These were found to have feature importance of 0.918, 0.014, and 0.068, where the (normalized) value measures the average reduction in training loss brought by the feature across all regression trees. For completeness, we considered the other variables of SnV$^-$ waveguide location, tip amplitude i.e. brightness of SnV$^-$ scatter from the waveguide tip, and FIB cut of the SnV$^-$ waveguide, but these variables did not result in training loss reduction in the current dataset.

Following the $\Delta ZPL_i$ prediction from XGBoost, we pass the predicted value to the COMSOL MPhDT to determine the orientation and waveguide cross-sectional location using a simple regression step--that is, we find the orientation and cross-sectional position that minimizes the difference between the MPhDT modeled $\Delta ZPL_i$ and the XGBoost-predicted $\Delta ZPL_i$. This assumes that the MPhDT is accurate enough to capture the strain in the QMC waveguide and is thus closer to the ground truth than the XGBoost model, and it is the same assumption used to determine SnV$^-$ strain susceptibilities using the EM algorithm. We note that future work could implement a feedback loop that \textit{progressively adjusts} the MPhDT to capture measured/predicted $\Delta ZPL_i$ that fall outside the numerically defined possible $\Delta ZPL_i$ range rather than \textit{enforcing} digitally defined bounds on the measured-modeled data fit. For this work, however, we deemed this as computationally expensive for FEM modeling and outside the scope of this research.

\section{Experimental Setup of Optical and Microwave Control in a Bluefors Cryostat}\label{sm:exp-bf}

The optical excitation and collection setups addressing our Bluefors cryostat are shown in Fig.~\ref{fig:optics_diagram}ab. The microwave excitation setup is shown in Fig.~\ref{fig:optics_diagram}c.

\begin{figure}
    \centering
    \includegraphics[width=\linewidth]{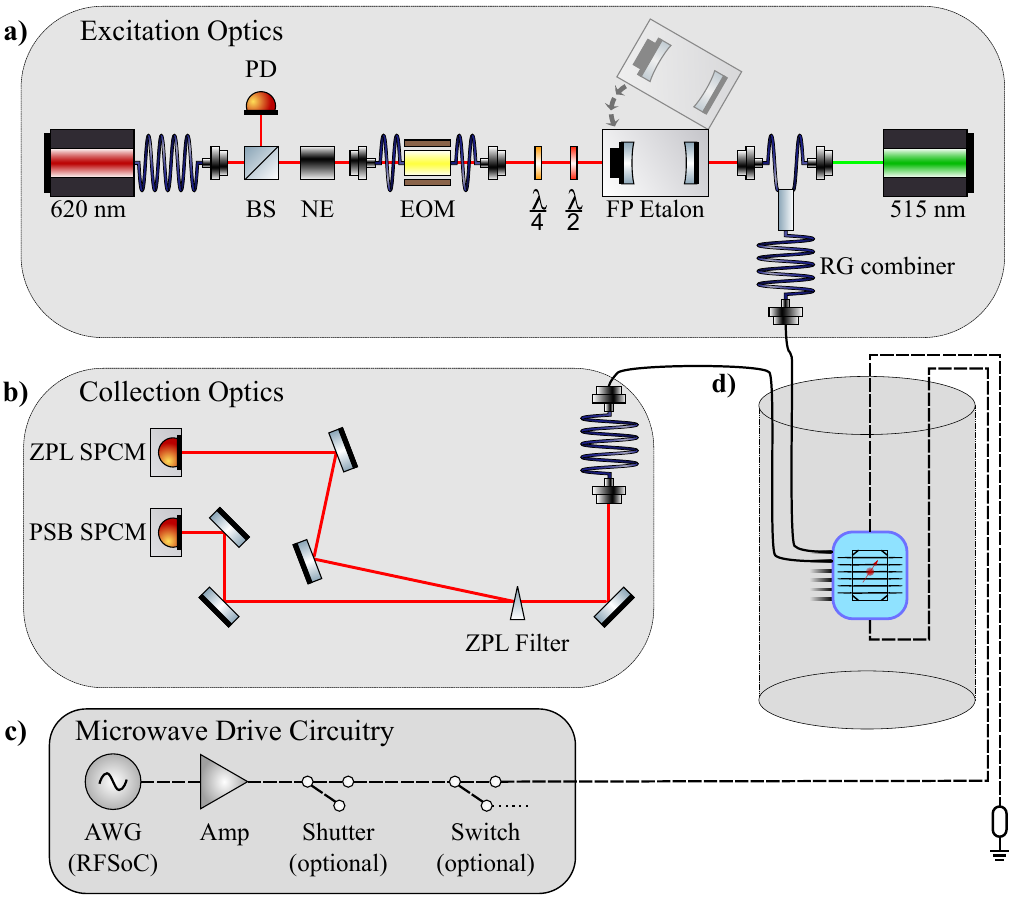}
    \caption{Experimental setup for sample loaded in a Bluefors LD250 cryostat. (a) excitation optics combining a resonant and repump laser in one optical channel of the sample. (b) collection optics separating the ZPL and PSB signals from the coupled channel of the sample. (c) Microwave circuit addressing the transmission line of the sample. (d) Optical and microwave fibers interfacing the cryostat.}
    \label{fig:optics_diagram}
\end{figure}

\section{Electromechanically Strain-Tuned Photoluminescence Excitation}\label{sm:strain-tuning}

\begin{figure}
    \centering
    \includegraphics[width=\linewidth]{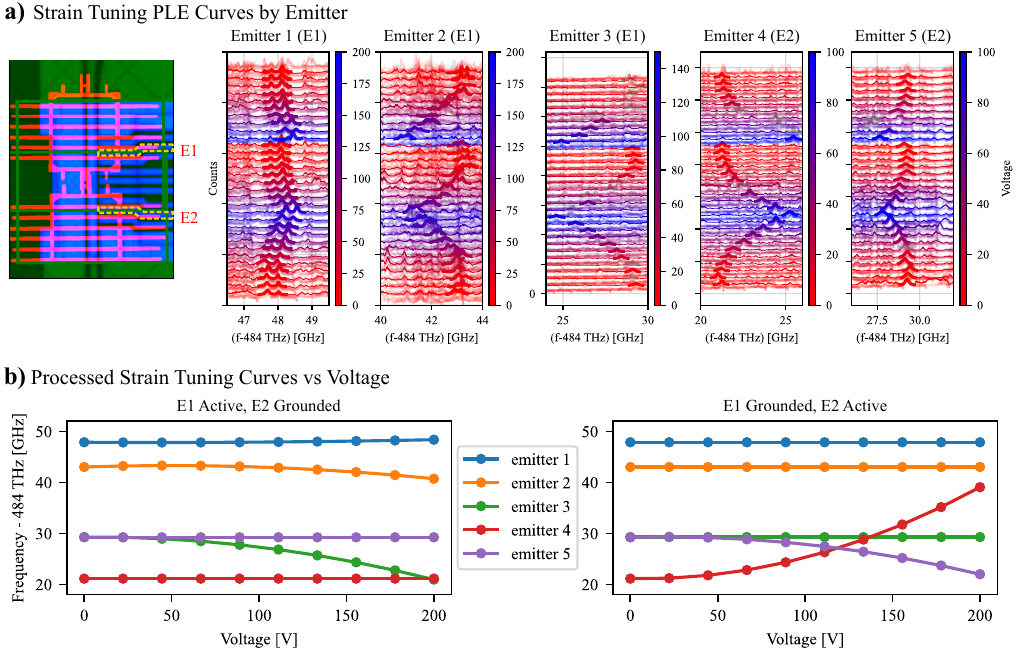}
    \caption{(a) Strain tuning curves for identified emitters. Emitters 1-3 are from the top pair of channels in the QR-PIC (tuned with electrode E1), and emitters 4-5 are from the bottom pair (tuned with electrode E2). Note that Emitter 1 is the SnV$^-$ identified in the main text. (b) processed frequency curves overlapped in space, indicating three ZPL curves that overlap one another. The first plot shows the curves under E1 applied voltage (E2-tuned emitters remain constant) and the second plot shows curves under E2 applied voltage (E1-tuned emitters remain constant).}
    \label{fig:SI_strain_tuned_ple}
\end{figure}

We probed the sample deployed in the Bluefors LD250 cryostat for SnV$^-$ emitters that exhibited ZPL tuning as follows. First, we perform a series of PLE experiments overlapping in frequency space under zero voltage, following the state machine result from \SMsec{sm:pulseseq}, to determine initial locations of $SnV^-$ ZPLs. We then increased the voltage applied to two select electrodes (labeled E1 and E2 in Fig.~\ref{fig:SI_strain_tuned_ple}a (left)) and repeated the PLE series at each voltage. The voltage sweep was performed by linearly increasing voltage from 0 to $V_{max}$, linearly decreasing back to 0, then stepping to $V_{max}$ and linearly decreasing to 0 again so as to check repeatability of any ZPL tuning detected. Fig.~\ref{fig:SI_strain_tuned_ple}a (right) shows the electromechanical strain tuning curves of five emitters. We fitted the ZPL tuning curves by seeding a curve fitting algorithm with an initial frequency guess (i.e. 484.048 THz as in Emitter 1) and using scipy's curve\_fit function to fit a double Lorentzian (Emitters 1-2) or a single Lorentzian (Emitters 3-5), dependent on the active status of the DC magnetic field generating Zeeman splitting during the PLE sweep. Upon fitting the Lorentzian for the initial PLE experiment at 0 V, we stored the center frequency as the seed guess for the subsequent PLE sweeps at each voltage index. After a first pass through curve fitting at all voltage indices, we fit the center frequencies to a second order polynomial mapping the ZPL frequency against voltage. Finally, we repeat the curve fitting using the second order polynomial frequency at each voltage value as the seed guess to refine the ZPL vs voltage curve fitting procedure. This double pass approach allowed us to detect low-amplitude peaks in the PLE data, which may be the result of poor beta factor coupling of an emitter to the QMC waveguide.

Fig.~\ref{fig:SI_strain_tuned_ple}b shows the projected ZPL curves of Emitters 1-5 as a function of applied voltage on E1 (left) and E2 (right). As the electrodes are spatially distant from one another with grounded electrodes in between, there is negligible projected crosstalk in the ZPL tuning curves, permitting selective tunability of ZPLs in the QR-PIC. We note that Emitters 3, 4, and 5 form a joint chain of ZPLs due to the ability of Emitter 4 to tune over the frequency of Emitters 3 and 5. By using the simplified expression for Hong-Ou-Mandel visibility,
\begin{equation}
    V_{HOM} = \frac{4\gamma_1\gamma_2}{(\gamma_1 + \gamma_2)^2 + (\nu_1 - \nu_2)^2},
\end{equation}
where $\gamma_i$ and $\nu_i$ are the linewidth and frequency of an emitter, respectively, we project the visibility matrix of the detected emitters in Table \ref{tab:visibility_table}.

\begin{table}[ht]
\centering
\begin{tabular}{c|ccccc}
    & 1     & 2     & 3     & 4     & 5 \\
    \hline
    1 (54.2 MHz) & ---   & \textcolor{gray}{0.850} & \textcolor{gray}{0.822} & \textcolor{orange}{0.868} & \textcolor{gray}{0.727} \\
    2 (122.7 MHz) &  & ---   & \textcolor{gray}{0.998} & \textcolor{orange}{0.999} & \textcolor{gray}{0.971} \\
    3 (133.2 MHz) &  &  & ---   & \textcolor{green}{0.995} & \textcolor{green}{0.983} \\
    4 (116.2 MHz) &  &  &  & ---   & \textcolor{green}{0.961} \\
    5 (173.0 MHz) &  &  &  &  & ---   \\
\end{tabular}
\caption{Projected Hong-Ou-Mandel visibilities between emitters documented in Fig.~\ref{fig:SI_strain_tuned_ple}. Green visibilities are reachable within 200 V applied to E1 or E2 and orange visibilities are reachable within 250 V. Gray visibilities are not reachable in the current device.}
\label{tab:visibility_table}
\end{table}

\section{Nuclear Spin Detection and Spin Memory Quantification via Electron Spin Echo}\label{sm:spin-echo}

\begin{figure}
    \centering
    \includegraphics[width=\linewidth]{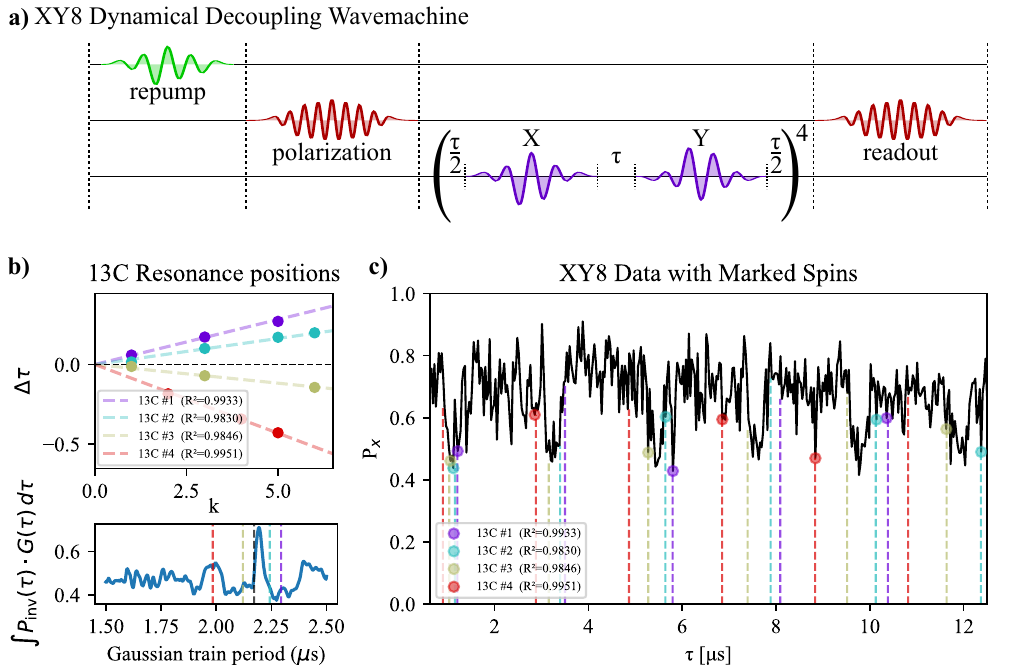}
    \caption{XY8 dynamical decoupling of the SnV$^-$ electron spin in this work. (a) wavemachine defined in \textit{Pulseseq}. (b) resonance positions of $^{13}\text{C}$ spins plotted as a fraction of the  $^{13}\text{C}$ Larmor period against $k$ order. These resonance positions are confirmed as peak signatures in the convolution of the XY8 spin echo data with a Gaussian pulse train. (c) XY8 spin data marked with each detected  $^{13}\text{C}$ spin.}
    \label{fig:SI_spin_detection}
\end{figure}

We applied a dynamical decoupling technique using an XY8 gate sequence to study the nuclear spin environment of the SnV$^-$ identified in the main text. The waveforms defining this decoupling sequence are shown in Fig.~\ref{fig:SI_spin_detection}a. To analyze spin signatures, we first found peaks in the XY8 decoupling data using scipy.signal's find\_peaks function. We then sorted the bins by $k$ order relative to the Larmor period for $^{13}C$ spins under 0.05 T magnetic field, $T_{L} = \frac{1}{\gamma_{^{13}C} B_0} = 1.868 \;\mathrm{\mu}\text{s}$. To identify peaks likely corresponding to a single $^{13}C$ spin, we fitted peaks to a linear curve $y = m k$; four such spins were identified with high r-squared value (Fig.~\ref{fig:SI_spin_detection}b top). We confirmed the spin signatures by convolving a Gaussian pulse train of period $\tau_v = \Delta\tau + T_L\left(k + \frac{1}{2}\right)$ with the XY8 decoupling data (Fig.~\ref{fig:SI_spin_detection}b bottom). 

The collapse troughs of identified spins are plotted overlaying the decoupling data in Fig.~\ref{fig:SI_spin_detection}c. These periodic signatures are used as initial guesses in an iterative algorithm to determine the full spin-echo signature of the electron spin, given by the form
\begin{equation}
    P(\tau) = b + c W(\tau) L(\tau),\;\ L(\tau)=\prod_{j=1}^NL_j(\tau; \gamma, B, A_{||_j}, A_{\perp_j}),
\end{equation}

where $W(\tau)$ is an incoherent envelope from a double-Lorentzian spin bath\cite{karapatzakis2024microwave}, $L_j(\tau)$ is the spin echo contribution of an individual nuclear spin ($\gamma$ is the gyromagnetic ratio of $^{13}\text{C}$, B is the magnetic field, and $A_{||}$ and $A_{\perp_j}$ are the spin-specific hyperfine coupling parameters), and $b$ and $c$ are the baseline and contrasts in the population, respectively. The iterative procedure is as follows:
\begin{enumerate}
    \item Fit a broad envelope of the form of a stretched exponential $W(\tau) = W_{0}(\tau) = \exp\left(\left(-\tau/T_2\right)^\xi\right)$ from spin echo data taken on a logarithmic scale from 1 $\mu\text{s}$ to 10 ms.
    \item Remove $W(\tau)$ from linear data taken in the 5-300 $\mu\text{s}$ regime. Fit, using a Maximum a Posteriori (MAP) least squares regression, a requested $N_{spins}$ number of nuclear spins interacting with the electron spin. Use a Metropolis sampling method to determine fitting uncertainties.
    \item While holding $L(\tau)$ fixed, fit the full $P(t)$ expression to the logarithmic scale spin echo data of varying pulse number XY-$N$, where $N \in \{1, 2, 4, 8, 16, 32\}$, using MAP and Metropolis sampling. Note that the $N=1$ case is a Hahn echo sequence and that, in this experiment, the $N=32$ case was an XY16 gate sequence repeated twice. Peak values determined by a moving maximum window are assigned a weight of 3 in fitting $W(\tau)$ to promote a stable solution capturing the envelope behavior and avoiding local minima in the fitting.
    \item Repeat steps 2 and 3 above for 5 iterations until convergence. Note that after the first step, $W(\tau)$ is determined using the full double-Lorentzian bath model.
\end{enumerate}

Fig.~\ref{fig:T2_analysis} shows the results of running the above algorithm for $1 \leq N_{spins} \leq 12$. We find that inputting $L(\tau)$ (Fig.~\ref{fig:T2_analysis}a demonstrating fitting for $N_{spins}=8$) in fitting $W(\tau)$ to the logarithmic data explains many collapse and revival signatures in the spin echo datasets (Fig.~\ref{fig:T2_analysis}b, $N_{spins}=8$), providing a clearer picture of the incoherent envelope decay of the model. We also fit an effective $T_2$ using the $1/e$ time of the envelope function $W(\tau)$ and find that the model converges to $T_2\sim1$ ms coherence times for 32 refocusing pulses (Fig.~\ref{fig:T2_analysis}d) and smaller fitting error (Fig.~\ref{fig:T2_analysis}e) as we approach $N_{spins} = 8$. We also note that the scaling exponent of $T_2$ vs pulse number of 0.644 approaches the expected 2/3 scaling \cite{wang2012comparison}. A scatter plot of fitted spin coupling parameters is shown in Fig.~\ref{fig:T2_analysis}f for all $N_{spins}$.

We note that the other collapse signatures in the spin echo data may result from (1) many spin signatures overlapping to create a broad collapse signature, (2) overlapping peaks that were not well-resolved in the current experiment due to time resolution limitations of the ZCU111 RFSoC and pulseseq package, or (3) interstitial $^{117}$Sn or $^{119}$Sn atoms unintentionally implanted in the QMC crystal during the ion implantation process. Even so, the current detection indicates the potential of the QR-PIC platform to control a nuclear spin memory bank for each waveguide-coupled SnV$^-$ \cite{taminiau2012detection, taminiau2014universal}.

\begin{figure}
    \centering
    \includegraphics[width=0.8\linewidth]{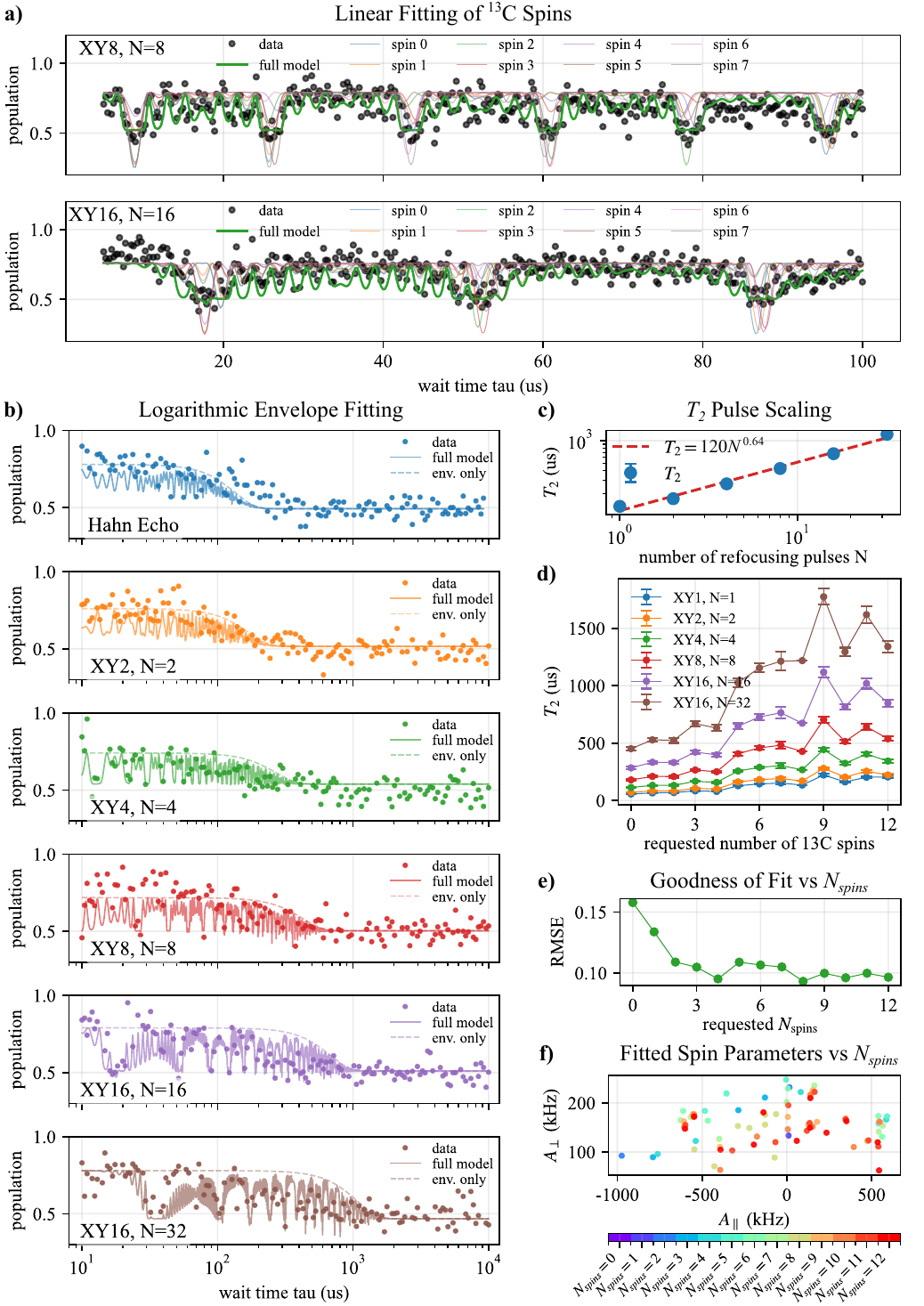}
    \caption{Spin echo analysis to quantify electron spin dephasing times. (a) fitting of $N_{spins}$ $^{13}\text{C}$ nuclear spins to dynamical decoupling data (XY8 and XY16) taken on a linear timescale. (b) Fitting a population function $P(\tau)$ to logarithmic data with varying number of refocusing pulses. (c) Plotting $T_2$ against refocusing pulse number. (d) Comparison of extracted $T_2$ for different requested number of $N_{spins}$. (e) Converging upon smaller root-mean-squared error (RMSE) with increasing $N_{spins}$. (f) Scatter plot of fitted coupling parameters of detected spins for different requested number of $N_{spins}$.}
    \label{fig:T2_analysis}
\end{figure}

\section{Exponential Scaling Prospects of QR-PICs}\label{sm:future-devices}

\begin{figure}
    \centering
    \includegraphics[width=\linewidth]{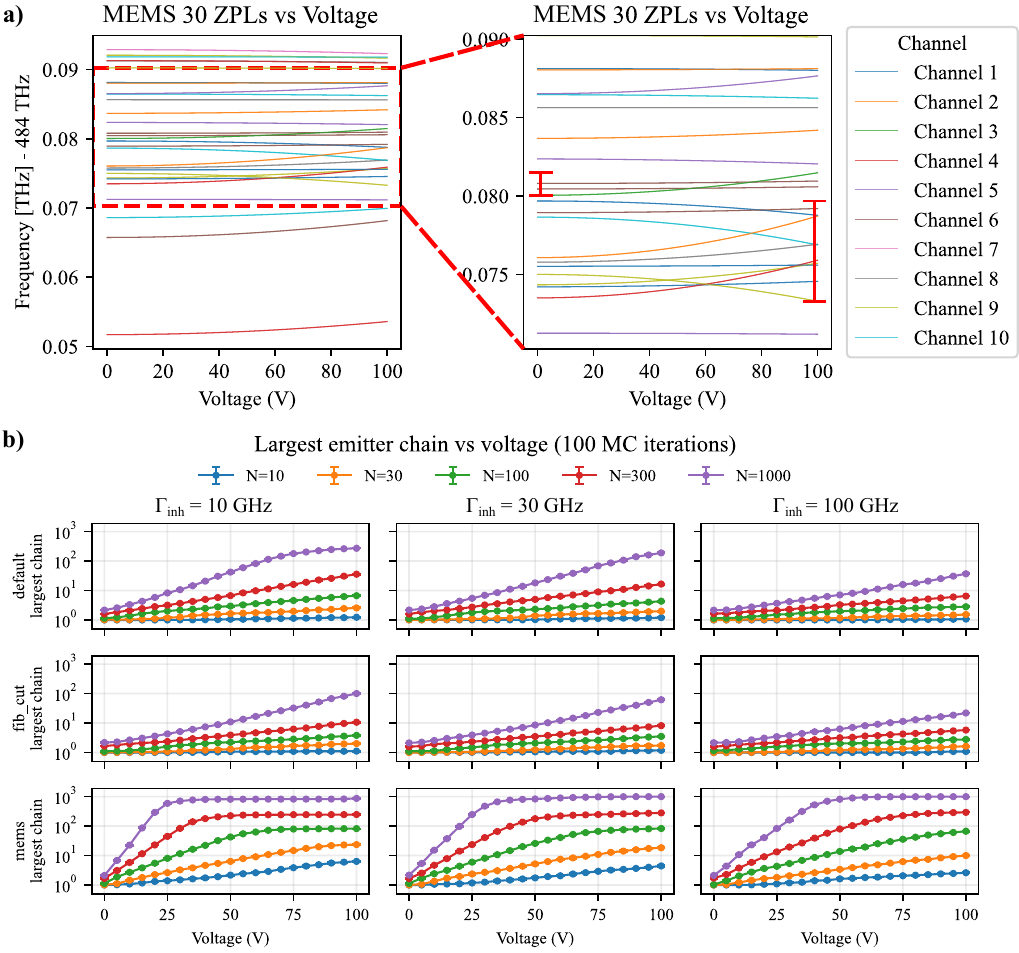}
    \caption{MPhDT simulation of 30 SnV$^-$ vacancy ZPLs in a QMC under applied strain tuning voltage. (a) shows the ZPL strain tuning curves overlayed with each other, with the inset indicating two portions of the graph that serve as individual connected chains due to overlapping curves. (b) plot of the largest connected chain in simulations of $n$ emitters under different QMC-$\Gamma_{inh}$ configurations as a function of voltage.}
    \label{fig:SI_exponential_scaling}
\end{figure}

We evaluated the exponential scalability of our QR-PIC platform by simulating, in COMSOL FEM, three configurations of microchiplets (defined in Fig.~\ref{fig:SI_COMSOL_FEM}) under electromechanical strain up to 100 V. 
For each microchiplet configuration, we randomly generate an ensemble of $N$ emitters. Each emitter has a random initial frequency $f_0$ sampled from an inhomogeneous distribution of linewidth $\Gamma_{inh} \in \{10, 30, 100\}$ GHz; a location between $-2\;\mu\text{m}$ and $2\;\mu\text{m}$ from the PIC origin, where the electromagnetic field generated by the microwave transmission line exhibits sufficient field for coherent electron spin control; and a random channel number from $1$ to $10$, where each channel is assumed to be independently tunable. 
For each emitter, we sample a location in the waveguide, with a sampling governed by the TRIM simulation distribution in $z$ and a weak Gaussian confinement in $y$. From each emitter's $xyz$ location, we extract ZPL tuning curves $\Delta f(V) = \alpha_2 V^2 + \alpha_1 V + \alpha_0$, such that the ZPL spectral location of an emitter $i$ is given by $f_{ZPL_i}(V) = \alpha_{2_i} V^2 + \alpha_{1_i} V + \left(f_{0_i}+\alpha_{0_i}\right)$.

Now, for each QMC and $\Gamma_{inh}$ combination, we performed a disjoint set union algorithm to determine the largest potential emitter chain in the simulation. This was done by considering pairs of emitters in different channels; if for some voltage $V_i$ $f_{min_j} < f_i(V_i) < f_{max_j}$ between emitters $i$ and $j$ (arbitrary indexing), then the emitters are considered "linkable" via a Bell Pair entanglement experiment. Then for a set of emitters $i \in \{1, ..., N\}$, the full set is linked if for every $i$ there is an emitter $j \in \{1, ..., N\}$, $i\neq j$ such that the frequency condition $f_{min_j} < f_i(V_i) < f_{max_j}$ is satisfied for some $V_i$.

Fig.~\ref{fig:SI_exponential_scaling} shows the plot of longest linker chain size versus maximum tuning voltage for each QMC and $\Gamma_{inh}$ combination. This informs us of (1) the most impactful modifications to the QR-PIC for the next generation of devices and (2) the electronic requirements by entangled system size. From this simulation, we find that the MEMS modifications of the QMC would most impact the QR-PIC performance, which aligns with an analytical estimation of longest linked chain size. 
For this estimation, we may assume that each emitter $i$ has a linearly tunable frequency--ignoring the quadratic tuning term--of
\begin{equation}
    f_{ZPL,i}(V) = f_{pre,i} + \Delta ZPL_i(V) = f_{pre,i} + \Delta ZPL_i + \left(\frac{df}{dV}\right)_i V
\end{equation}
If we assume a normal distribution for each per-emitter variable--$f_{pre,i}\sim\mathcal{N}(\mu_0,\sigma_{inh}^2)$ of inhomogeneous frequencies of SnV$^-$ with standard deviation $\sigma_{inh}$, frequency shifts $\Delta ZPL_i\sim\mathcal{N}(0,\sigma_\Delta^2)$, and strain tuning per unit voltage $\left(\frac{df}{dV}\right)_i \sim \mathcal{N}(\mu_s,\sigma_s^2)$--with covariances $\text{Cov}(f_{pre},\Delta ZPL) = C_1$ and $\text{Cov}\left(\Delta ZPL, \frac{df}{dV}\right) = C_2$. Each emitter sweeps a frequency interval 
\begin{equation}
\mathcal{I}_i = \left[f_i + s_i \cdot V_{\min},\; f_i + s_{i} \cdot V_{\max}\right],
\end{equation}
where $f_i = f_{pre,i} + \Delta ZPL_i$ and $s_i =  \left(\frac{df}{dV}\right)_i$. This gives the width and center of each frequency interval as 
\begin{equation}
    w_i = \abs{s_i}\left(V_{\max} - V_{\min}\right) = \abs{s_i}\delta V,\;c_i = f_i + s_i \cdot \frac{V_{\min} + V_{\max}}{2} = f_i + s_i \bar{V}.
\end{equation}
The variance of the center frequency, given our assumed distributions and covariances, becomes
\begin{equation}
    \sigma_c^2 = \sigma_{inh}^2 + \sigma_\Delta^2 + \bar{V}^2\sigma_s^2 + 2 C_1 + \bar{V} C_2.
\end{equation}
We now consider the overlap condition and probability between emitters $i$ and $j$. Two emitters can be made indistinguishable if their tuning ranges overlap, or $\abs{c_i - c_j} \leq \left(w_i + w_j\right)/2$. The right hand side is the sum of the half-widths of each frequency interval. The probability of this occurring becomes
\begin{equation}
    p_{link} = P\left(\abs{c_i - c_j} \leq \left(w_i + w_j\right)/2\right) = \mathbb{E}_{w_i,w_j}\left[\left(2\Phi \frac{w_i+w_j}{2\sqrt{2}\sigma_c}\right)-1\right],
\end{equation}
where $\Phi$ is the cumulative distribution function.

All emitter frequency intervals together form an interval graph, where the largest chain, or maximum clique, is the maximum number of mutually overlapping intervals. To estimate this, we now consider the probability of an emitter's interval covering a frequency $f$:
\begin{align}
    p_i(f) &= \int_{-\infty}^{\infty}\mathcal{N}(s;\mu_s,\sigma_s^2)\int_{f-\abs{s}\frac{\delta V}{2}}^{f+\abs{s}\frac{\delta V}{2}}\mathcal{N}(c;\mu_c,\sigma_c^2)dc\;ds \\
    &= \int_{-\infty}^{\infty}\mathcal{N}(s;\mu_s,\sigma_s^2)\left[\Phi\left(\frac{f - \mu_c + \frac{\abs{s}\delta V}{2}}{\sigma_c}\right)-\Phi\left(\frac{f - \mu_c - \frac{\abs{s}\delta V}{2}}{\sigma_c}\right)\right]\;ds.
\end{align}

At the peak frequency $\mu_c$, the expected longest chain is therefore
\begin{equation}
    \langle n_{max} \rangle = N \int_{-\infty}^{\infty}\mathcal{N}(s;\mu_s,\sigma_s^2)\left[\Phi\left(\frac{\abs{s}\delta V}{2\sigma_c}\right)-\Phi\left(\frac{-\abs{s}\delta V}{2\sigma_c}\right)\right]ds.
\end{equation}

In a limit of nearly uniform tuning rates, this equation simplifies to $\langle n_{max} \rangle \approx N\left[2\Phi\left(\frac{\mu_s\delta V}{2\sigma_c}\right) - 1\right]$. This suggests, as our simulations roughly imply, that the longest chain grows linearly with $N$ and saturates as the tuning range $\mu_s\delta V$ exceeds the inhomogeneous frequency spread $\sigma_c$. In practice, due to varying emitter strain tunability and the quadratic nature of the spectral tuning curves versus voltage, the expected largest emitter chain on our systems must be numerically evaluated against the MPhDT.

\end{supplement}
\newpage
\printbibliography[heading=none]
\end{refsection}
\end{document}